\documentclass[longbibliography]{jpp}
\pdfoutput=1

\usepackage{graphicx}
\usepackage{natbib}
\usepackage{amsmath}
\usepackage{color}

\ifCUPmtlplainloaded \else
  \checkfont{eurm10}
  \iffontfound
    \IfFileExists{upmath.sty}
      {\typeout{^^JFound AMS Euler Roman fonts on the system,
                   using the 'upmath' package.^^J}%
       \usepackage{upmath}}
      {\typeout{^^JFound AMS Euler Roman fonts on the system, but you
                   dont seem to have the}%
       \typeout{'upmath' package installed. JPP.cls can take advantage
                 of these fonts, if you use 'upmath' package.^^J}%
      }
  \else
  \fi
\fi

\ifCUPmtlplainloaded \else
  \checkfont{msam10}
  \iffontfound
    \IfFileExists{amssymb.sty}
      {\typeout{^^JFound AMS Symbol fonts on the system, using the
                'amssymb' package.^^J}%
       \usepackage{amssymb}%
       \let\le=\leqslant  
       \let\ge=\geqslant  
      }{}
  \fi
\fi

\ifCUPmtlplainloaded \else
  \IfFileExists{amsbsy.sty}
    {\typeout{^^JFound the 'amsbsy' package on the system, using it.^^J}%
     \usepackage{amsbsy}}
    {}
\fi

\newcommand{\change}[1]{#1}
\newcommand{\changetwo}[1]{#1}

\newcommand\zhat{\hat{z}}

\newcommand{\Alfven}{Alfv\'{e}n }
\newcommand{\Alfvenic}{Alfv\'{e}nic }

\newcommand{\V}[1]{\mathbf{#1}} 

\title[Field-Particle Correlations \& Ion Cyclotron Turbulence]
      {Diagnosing collisionless energy transfer using field-particle correlations: \\Alfv\'en-Ion Cyclotron Turbulence}

\author[Klein and Others]%
{Kristopher G. Klein
\thanks{Email address
for correspondence: kgklein@lpl.arizona.edu}$^{1}$,
Gregory G. Howes$^{2}$, Jason M. TenBarge$^{3}$,
Francesco Valentini$^{4}$} 

\affiliation{$^1$ Lunar and Planetary Laboratory, University of Arizona, Tucson, AZ 85719, USA\\
[\affilskip]
$^2$Department of Physics and Astronomy, 
University of Iowa,
Iowa City, IA 52242, USA\\
[\affilskip]
$^3$Department of Astrophysical Sciences, Princeton University, Princeton, NJ 08544,
USA\\
[\affilskip]
$^4$ Dipartimento di Fisica, Universit\`a della Calabria, I-87036 Cosenza, Italy\\
}

\pubyear{2018}
\volume{?}
\pagerange{?}
\date{?; revised ?; accepted ?.}

\begin{document}

\maketitle

\begin{abstract}
We apply field-particle correlations--- a technique that tracks the
time-averaged velocity-space structure of the energy density transfer
rate between electromagnetic fields and plasma particles---to data
drawn from a hybrid Vlasov-Maxwell simulation of Alfv\'en
Ion-Cyclotron turbulence. Energy transfer in this system is expected
to include both Landau and cyclotron wave-particle resonances, unlike
previous systems to which the field-particle correlation technique has
been applied. In this simulation, the energy transfer rate mediated by
the parallel electric field $E_\parallel$ comprises approximately
$60\%$ of the total rate, with the remainder mediated by the
perpendicular electric field $E_\perp$. The parallel electric field
resonantly couples to protons, with the canonical bipolar
velocity-space signature of Landau damping identified at many points
throughout the simulation.  The energy transfer mediated by $E_\perp$
preferentially couples to particles with \changetwo{$v_{tp} \lesssim
  v_\perp \lesssim 3 v_{tp}$} in agreement with the expected formation
of a cyclotron diffusion plateau. Our results demonstrate clearly that
the field-particle correlation technique can distinguish distinct
channels of energy transfer using single-point measurements, even at
points in which multiple channels act simultaneously, and can be used
to determine quantitatively the rates of particle energization in each
channel.
\end{abstract}

\section{Introduction}
\label{sec:intro}

Identifying the mechanisms that transport energy between
electromagnetic fields and charged particles in nearly collisionless
plasmas is a critical step in the broader effort to characterize and
ultimately predict the dissipation of turbulence in space and
astrophysical plasmas.  Proposed mechanisms for energy transfer can
broadly be grouped into three classes: (i) resonant mechanisms,
\emph{e.g.}, Landau damping, Barnes damping, or cyclotron damping
\citep{Landau:1946,Barnes:1966,Kennel:1966}; (ii) non-resonant
mechanisms, \emph{e.g.}, stochastic heating by low-frequency,
large-amplitude kinetic \Alfven waves
\citep{McChesney:1987,Chen:2001,Johnson:2001,Chandran:2010a,Chandran:2010b}
or magnetic pumping \citep{Berger:1958,Lichko:2017}; and (iii) spatially localized
mechanisms, \emph{e.g.}, magnetic reconnection at intermittent current
sheets
\citep{Dmitruk:2004,Matthaeus:2011,Servidio:2011a,Karimabadi:2013,Zhdankin:2013,Osman:2014a,Osman:2014b,Zhdankin:2015a}.
The solar wind, a hot and diffuse plasma emanating from the Sun,
serves as a natural laboratory for observing which energization
mechanisms operate under what plasma conditions. A significant
limitation of \emph{in situ} measurements of the solar wind is that
most observations occur at a single point, therefore it is not
possible to assess the entire energy budget of the system.  However,
as different mechanisms preferentially transfer energy to particles
with specific characteristic velocities, single-point observations of
the velocity-space structure of the energy transfer may enable the
determination of which energization mechanisms are at work.

A field-particle correlation technique \citep{Klein:2016,Howes:2017a}
has been proposed to capture the velocity-space structure of
energization mechanisms from single-point observations. This technique
resolves the electric-field component of the field-particle
interaction term in the Vlasov equation as a function of velocity and
averages the energy density transfer rate over some correlation time
interval.  By capturing the transfer rate as a function of velocity,
the regions in phase space that lose energy to or gain energy from the
fields are identified. Performing a time average removes the
oscillatory energy transfer between the plasma and the fields,
isolating the secular component of the transfer that leads to net
energization. Combined, this velocity-resolved and time-averaged
transfer rate, denoted the \emph{velocity-space signature}, can be
used to characterize the energization mechanisms operating in a plasma
measured only at a single point in space.

Previous applications of this field-particle correlation technique
include numerical studies of electrostatic waves
\citep{Klein:2016,Howes:2017a} and instabilities \citep{Klein:2017a},
monochromatic kinetic \Alfven waves \citep{Howes:2017b}, energization
near current sheets arising from strong \Alfven wave collisions
\citep{Howes:2018a}, as well as low-frequency, wavevector anisotropic,
strong turbulence \citep{Klein:2017b}. The technique has also been
applied to turbulent magnetosheath plasma measured by MMS
\citep{Chen:2019}. For both simulations and observations, a clear
signature of energy transfer as a function of $v_\parallel$ was
identified, which is indicative of significant energy being
transferred via the Landau resonance. The previous numerical
simulations of turbulence used \texttt{AstroGK}, a gyrokinetic code in
which the low-frequency approximation arising through the
gyroaveraging procedure eliminates the physics of the cyclotron
resonance \citep{Howes:2006}.  In this work, we use a hybrid
Vlasov-Maxwell code, \texttt{HVM}, to simulate higher frequency
Alfv\'en-ion cyclotron turbulence, a system in which proton cyclotron
damping may contribute to the removal of energy from the
turbulence. For the Alfv\'en-ion cyclotron system, both $E_\parallel$
and $E_\perp$ may contribute to the energy density transfer via the
Landau and cyclotron resonances, respectively.  At most points
throughout the simulation, resonant signatures near the proton thermal
velocity, $|v_\parallel| \sim v_{tp}$, are associated with
energization due to $E_\parallel$, while particles with $v_{tp}
\lesssim v_\perp \lesssim 3 v_{tp}$ couple most strongly with
$E_\perp$.  By diagnosing the energy transfer at 64 spatial points
distributed throughout the simulation, we find that the energy
transfer mediated by $E_\parallel$ \changetwo{after one Alfv\'en
  crossing time} at these points accounts for \changetwo{$62\% \pm
  24\%$ of the total energy transfer.}

The remainder of this paper is organized as follows.  An overview of
the relevant damping mechanisms and the simulation code employed,
\texttt{HVM}, is given in Secs.~\ref{sec:mech} and \ref{sec:hvm}.
The field-particle correlation method is presented in
Sec.~\ref{sec:fpem} and is applied to simulation data in
Sec.~\ref{sec:vel}.  In Sec.~\ref{sec:heatflux}, we discuss the
relative importance of the electric field and advection to energy
transfer, followed by conclusions in Sec.~\ref{sec:conclude}.  This
extension of the field-particle correlation technique to a regime of
higher-frequency turbulence, distinct from previous numerical studies
of low-frequency turbulence, demonstrates that this technique can
successfully employ single-point measurements both to distinguish
distinct mechanisms of energy transfer and to determine quantitatively
the rates of particle energization in each channel.


\section{Energy Transfer in Ion-Cyclotron Turbulence}
\label{sec:mech}

Collisionless resonant mechanisms that mediate energy transfer in
magnetized plasmas sensitively depend on the frequency of the
associated plasma fluctuations.  These mechanisms require a portion of
the particle velocity distribution with significant phase space
density to approximately satisfy the resonance condition
$\omega(\V{k}) - k_\parallel v_\parallel - n\Omega_s = 0$, where
$\omega(\V{k})$ is the wavevector dependent normal mode frequency,
$k_\parallel$ is the component of the wavevector parallel to the mean
magnetic field $\V{B}_0$, $v_\parallel$ is the parallel particle
velocity, $\Omega_s = q_s B/m_s c$ is the cyclotron frequency for
species $s$, and $n$ is an integer. Previous field-particle
correlation work specifically focused on energy transfer in systems
where the Landau, or $n=0$, resonance is the only available channel
for collisionless damping, including both systems with monochromatic
waves \citep{Klein:2016,Howes:2017a,Klein:2017a,Howes:2017b} and
simulations of strong, wavevector-anisotropic turbulence
\citep{Klein:2017b,Howes:2018a}.

The Landau resonance is important for low-frequency, wavevector
anisotropic fluctuations of the kind typically observed in the solar
wind. A significant body of evidence, including observational
\citep{Sahraoui:2010a,Chen:2013a,Roberts:2015b}, theoretical
\citep{Schekochihin:2009,Kunz:2015,Kunz:2018}, and numerical
\citep{Howes:2008a,Mallet:2015,Groselj:2018} studies, suggests that
magnetized collisionless turbulence is dominated by low-frequency,
anisotropic \Alfvenic fluctuations. However, as discussed in
\cite{Cerri:2016} and \cite{Arzamasskiy:2019}, the role of
higher-frequency fluctuations in realistic turbulent systems is
still an area of active debate.  For higher frequency fluctuations,
with turbulent fluctuation frequencies at or above the proton
cyclotron frequency $\omega \gtrsim \Omega_p$, collisionless damping
may proceed through the $n \ne 0$ cyclotron resonances.

\begin{figure}
  \centerline{\includegraphics[width = 14cm, viewport = 0 30 300 300, clip=true]
    {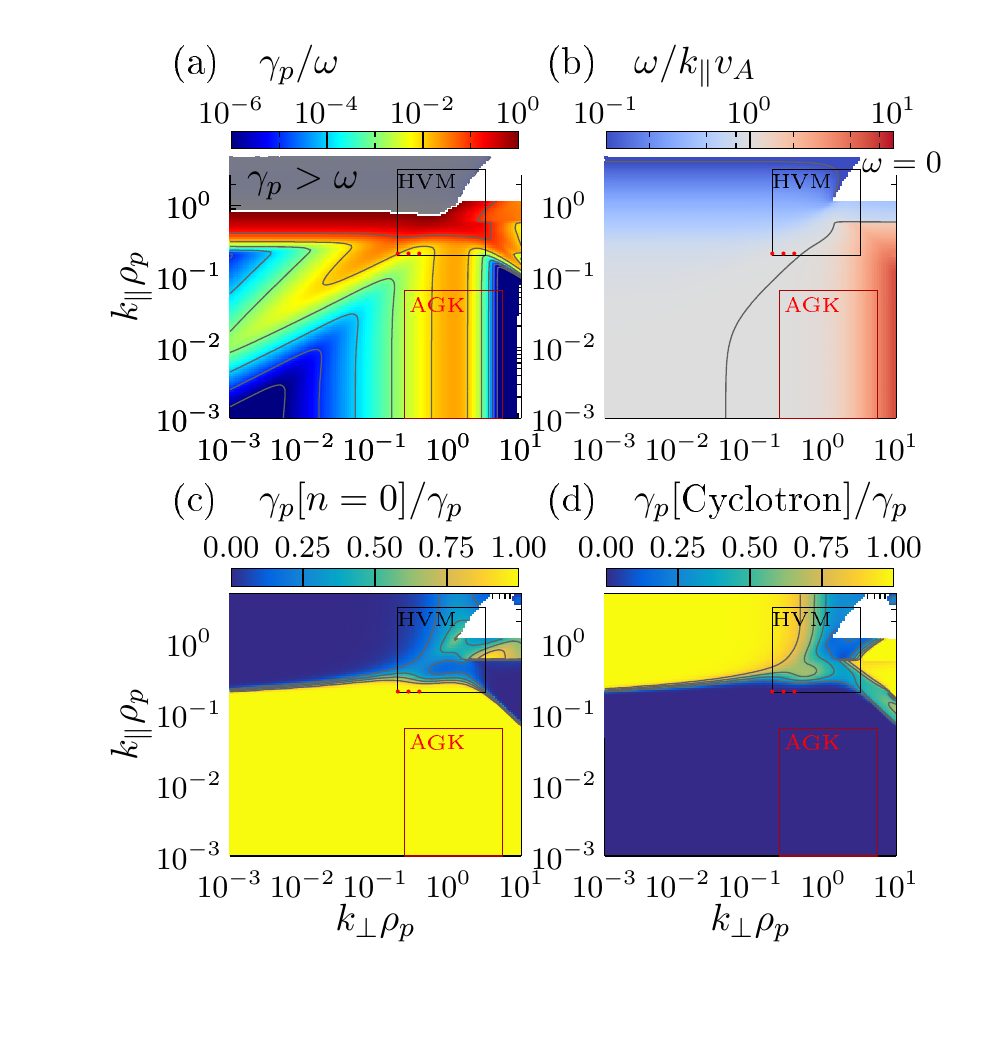}}%
  \caption{\change{Eigenfunction relations} for the \Alfven dispersion
    surface as a function of $\V{k} \rho_p$ for a $\beta_p=1$ plasma
    (in which $\rho_p=d_p$). (a) The normalized total proton damping
    rate $\gamma_p/\omega$ from Eqn.~\ref{eqn:power}. (b) The
    normalized parallel phase velocity $\omega/k_\parallel v_A$. (c)
    The fraction of the proton damping rate due to the Landau
    resonance. (d) The fraction of the proton damping rate due to the
    cyclotron resonance. The boxes outline the wavevector ranges for
    \texttt{HVM} simulations presented here (black) and in previous
    gyrokinetic simulations of low-frequency, strong turbulence (red)
    \cite{Klein:2017b}. \change{The red dots indicate the values of
      $(|k_\perp|,|k_\parallel|)\rho_p$ with initialized \Alfven waves
      for the HVM simulation. The grey region in the upper left-hand
      corner shows where $\gamma_p > \omega$, and the white region in
      the upper right-hand corner shows where $\omega=0$. }}
\label{fig:linear}
\end{figure}


In this work, we focus on determining the velocity-space
signatures of energy transfer to the protons in higher frequency,
Alfv\'en-Ion Cyclotron turbulence. In order to select a wavevector
region for which cyclotron damping may be present, we consider the
collisionless power absorption for the \Alfven dispersion surface as
derived from linear kinetic theory. The power absorption by species $s$ due to
a normal mode with frequency $\omega(\V{k})$ in one wave period,
following \cite{Quataert:1998}, is given by
\begin{equation}
\frac{\gamma_s(\V{k})}{\omega(\V{k})} = \frac{\V{E}^*(\V{k})\cdot
  \underline{\underline{\Lambda}}_s^a(\V{k}) \cdot \V{E}(\V{k})}{4 W_{\rm EM}(\V{k})}.
\label{eqn:power}
\end{equation}
The Fourier-transformed vector electric field and its complex
conjugate are given by $\V{E}(\V{k})$ and $\V{E}^*(\V{k})$, the
electromagnetic wave energy by $W_{\rm EM}(\V{k})$ and the
anti-Hermitian part of the linear susceptibility tensor for species
$s$ is $\underline{\underline{\Lambda}}_s^a(\V{k})$. The decomposition
of the power absorption by species given by Eqn.~\ref{eqn:power} is
valid as long as the total damping rate is small compared to the wave
frequency $\sum_s\gamma_s < \omega$. In Figure~\ref{fig:linear}(a), we
use Eqn.~\ref{eqn:power} to compute the proton power absorption for
the \Alfvenic dispersion surface for a proton-electron plasma with
$\beta_p =8 \pi n_p T_p/B^2= 1$ and $T_p = T_e$ calculated using the
PLUME dispersion solver \citep{Klein:2015a}, showing significant
proton damping primarily in two regions\footnote{The white triangle
  for $k_\perp \rho_p > k_\parallel \rho_p > 1$ represents the
  wavevector region where the \Alfven mode is non-propagating with
  $\omega=0$, causing Eqn.~\ref{eqn:power} to be invalid.}: (i)
$k_\perp \rho_p \sim 1$ (yellow) and (ii) $k_\parallel \rho_p \gtrsim
1$ (red\footnote{\change{The region where $\gamma_p>\omega$, and thus
    linear theory is formally invalid for the Alfv\'en solution, is
    shaded in grey.}}).  The parallel wave phase velocity
$\omega/k_\parallel v_A$ is plotted in Figure~\ref{fig:linear}(b),
showing three general regimes: (i) the non-dispersive \emph{MHD
  \Alfven wave regime} with $k_\parallel \rho_p \ll 1$ and $k_\perp
\rho_p <1 $ where $\omega/k_\parallel v_A=1$; (ii) the \emph{ion
  cyclotron wave regime} with $k_\parallel \rho_p \gtrsim 1$ where the
phase velocity decreases as $k_\parallel \rho_p$ increases; and (iii)
the \emph{kinetic \Alfven wave regime} with $k_\parallel \rho_p \ll 1$
and $k_\perp \rho_p \gtrsim 1 $ where the phase velocity increases as
$k_\perp \rho_p$ increases.  Note that, for a plasma with $\beta_p=1$,
the proton Larmor radius $\rho_p=v_{tp}/\Omega_p$ is the same as the
proton inertial length $d_p=v_A/\Omega_p$, as the scales can be
related via $\rho_p = d_p/\sqrt{\beta_p}$.

To quantify the relative contributions to the proton damping rate
$\gamma_p$ from Landau and cyclotron damping, we recalculate
Eqn.~\ref{eqn:power} using a susceptibility tensor
$\underline{\underline{\Lambda}}_p$ constructed using only the $n=0$
contributions (Landau damping) or $n \neq 0$ contributions to the
$(x,y)$ manifold (cyclotron damping, c.f. \cite{Stix:1992} $\S 11.8$).
This decomposition by the characteristic resonance shows that the two
primary regions of significant proton damping are caused by distinct
mechanisms.  In Figure~\ref{fig:linear}(c), we plot the ratio of the
Landau damping rate to the total proton damping rate
$\gamma_p[n=0]/\gamma_p$, showing that, in the region $k_\parallel \rho_p
\ll 1$, Landau damping is dominant, so that the yellow region at $k_\perp
\rho_p \sim 1$ and $k_\parallel \rho_p \ll 1$ in
Figure~\ref{fig:linear}(a) is dominated by Landau damping.  In
Figure~\ref{fig:linear}(d), we plot the ratio of the cyclotron damping
rate to the total proton damping rate
$\gamma_p[\mbox{cyclotron}]/\gamma_p$, showing that, in the region
$k_\parallel \rho_p \gtrsim 1$, cyclotron damping is dominant, so the
red and black regions at $k_\parallel \rho_p \gtrsim 1$ in
Figure~\ref{fig:linear}(a) is dominated by cyclotron damping.




For Landau damping of \Alfven waves in the wavevector anisotropic
region with $k_\perp\rho_p \sim 1$ and $k_\parallel \ll k_\perp$, the
collisionless energy transfer is associated with resonant parallel
phase velocities $\omega/k_\parallel \sim v_{A}$, which are of order
$v_{tp}$ for plasmas with $\beta_p \approx 1$.  For waves with
$k_\perp \rho_p \gg 1$, the parallel phase velocity of the wave
increases, moving out of resonance with the thermal proton population,
reducing the effectiveness of proton Landau damping. As the parallel
wavevector $k_\parallel \rho_p$ increases to unity and beyond, the
parallel phase velocity decreases $\omega/k_\parallel \rightarrow 0$,
similarly leading to a quenching of Landau damping.

For cyclotron damping, the velocity distribution evolves along
circular pitch angle contours centered about the parallel wave phase
velocity, where this pitch-angle diffusion drives the distribution
toward a state where it is constant along contours
$(v_\parallel-\omega/k_\parallel)^2+v_\perp^2$
\citep{Kennel:1966,Marsch:2001,He:2015}. For a spectrum of proton
cyclotron waves propagating both up and down the magnetic field, with
$k_\parallel >0$ and $k_\parallel<0$, this evolution leads to the
formation of a quasilinear \emph{cyclotron diffusion plateau} in the
region with significant overlap of constant energy contours with
$v_{tp} \lesssim v_\perp \lesssim 3 v_{tp}$. \changetwo{The parallel
  structure of this plateau peaks at small $v_\parallel$,
  corresponding to higher phase-space densities near the center of the
  proton distribution.}

With the identification of the different regions of wavevector space
$(k_\perp \rho_p, k_\parallel \rho_p)$ in which Landau or cyclotron
damping are expected to dominate, as shown in Figure~\ref{fig:linear},
we may now specify an appropriate wavevector range to yield
significant proton cyclotron damping in a simulation of high-frequency
Alfv\'en-ion cyclotron turbulence.


\section{Hybrid Simulations of Alfv\'en-Ion Cyclotron Turbulence}
\label{sec:hvm}

Based upon these power absorption calculations, we select a wavevector
region for which both Landau and cyclotron damping may be active. For
our \texttt{HVM} simulation of Alfv\'en-ion cyclotron turbulence, we
simulate a turbulent plasma in a domain over a wavevector range $0.2
\le k_\perp d_p \le 3.2$ and $0.2 \le k_\parallel d_p \le 3.2$,
denoted by the black box in Figure~\ref{fig:linear}. For comparison,
the previous turbulent gyrokinetic simulations used in
\cite{Klein:2017b} spanned $0.25 \le k_\perp d_p \le 5.5$ under the
asymptotic anisotropic conditions $k_\parallel \ll k_\perp$ of the
gyrokinetic approximation, a wavevector range denoted by the red box
in Figure~\ref{fig:linear}.  To describe turbulent fluctuations with
finite parallel wavevectors $k_\parallel d_p \gtrsim 1$ and
ion-cyclotron frequencies $\omega \sim \Omega_p$, we employ the hybrid
Vlasov-Maxwell code \texttt{HVM} \citep{Valentini:2007}. \texttt{HVM}
self-consistently solves the Vlasov equation for ions on a uniform
fixed 3D grid in physical space and a uniform fixed 3V grid in
velocity space, coupled with an isothermal fluid description for the
electrons through Maxwell's equations. This method allows for accurate
simulation of ion kinetic-scale phenomena. By employing an Eulerian
approach, these simulations are able to resolve velocity-space
structure without the statistical noise associated with
particle-in-cell macroparticles. Since the ions are fully kinetic, we
resolve ion-cyclotron frequency physics, which is outside the
gyrokinetic formalism.

The simulation employs $32^3$ spatial grid points and $51^3$ velocity
grid points. The velocity grid spans $\pm 5 v_{tp}$ for all three
directions, and the size of the isotropic simulation cube is $L=10 \pi
d_p$.  The proton plasma beta is unity, $\beta_p=1$, and the proton
and electron temperatures are in equilibrium $T_p = T_e$. The uniform
background magnetic field is in the $\hat{\V{z}}$ direction,
$\V{B}_0=B_0 \zhat$.  \change{The simulation dissipates small scale
  fluctuations using grid-scale resistivity by adding an $\eta J$ term
  into Ohm's law.  A small value for the resistivity $\eta$ has been
  chosen in order to achieve relatively high Reynolds numbers and to
  remove any spurious numerical effects due to the presence of grid-scale
  current sheets. The choice of this small value for the resistivity
  corresponds to a very small correction, confined to small scales,
  with the resulting dissipation electric field $\eta J$ only becoming
  dominant for largest wave numbers in simulation.}

Twelve \Alfven wave modes at the largest two spatial scales in the
domain are initialized: $\V{k}d_p=(k_x d_p, k_y d_p, k_z d_p)=
(0.2,0,\pm0.2),$ $(0,0.2,\pm0.2),$ $(0.2,0.2,\pm0.2),$
$(-0.2,0.2,\pm0.2),$ $(0.4,0,\pm0.2),$ and $(0,0.4,\pm0.2)$. The
magnetic and velocity fluctuations satisfy the MHD \Alfven wave
eigenfunctions and are assigned distinct random phases $\phi_k \in
[0,2\pi]$ for each initialized wavevector $\V{k}$. The real amplitude
of each Fourier wavevector mode is chosen so that the system will have
a sufficiently strong turbulent cascade, as measured by the
nonlinearity parameter, $\chi = (k_\perp/k_\parallel) (\delta B_\perp/
B_0) \approx 1$; we set amplitudes $ \delta \hat{B}_k = 1/\sqrt{2}$
for $\V{k}d_p = (0.2,0,\pm0.2)$ and $(0.0,0.2,\pm0.2)$, $ \delta
\hat{B}_k = 1/4$ for $\V{k}d_p = (0.2,0.2,\pm0.2)$ and
$(-0.2,0.2,\pm0.2)$, and $ \delta \hat{B}_k = 1/(4\sqrt{2})$ for $
\V{k}d_p =(0.4,0.0,\pm0.2)$ and $(0.0,0.4,\pm0.2)$, which corresponds
to an overall initial RMS amplitude of $\delta B_\perp/ B_0=1/2$. In
contrast to gyrokinetic simulations, where the significant wavevector
anisotropy $k_\parallel \ll k_\perp$ allows the turbulence to be
strong (\emph{i.e},. $\chi \sim 1$) for $\delta B_\perp/B_0 \ll 1$,
having a system of strong turbulence for the wavevectors considered
here with $k_\parallel \sim k_\perp$ requires $\delta B_\perp \sim
B_0$.

\changetwo{This simulation box size was intentionally chosen to
  enclose wavevectors susceptible to both Landau and cyclotron
  resonances, allowing the application of the field-particle
  correlation technique to systems in which multiple heating
  mechanisms operate.  This work does not necessarily replicate solar
  wind turbulence, which is typically found to have more significant
  wavevector anisotropies than are simulated here, as described for
  instance in \cite{Chen:2016a}.}

\begin{figure}
  \centerline{\includegraphics[width = 14cm]
    {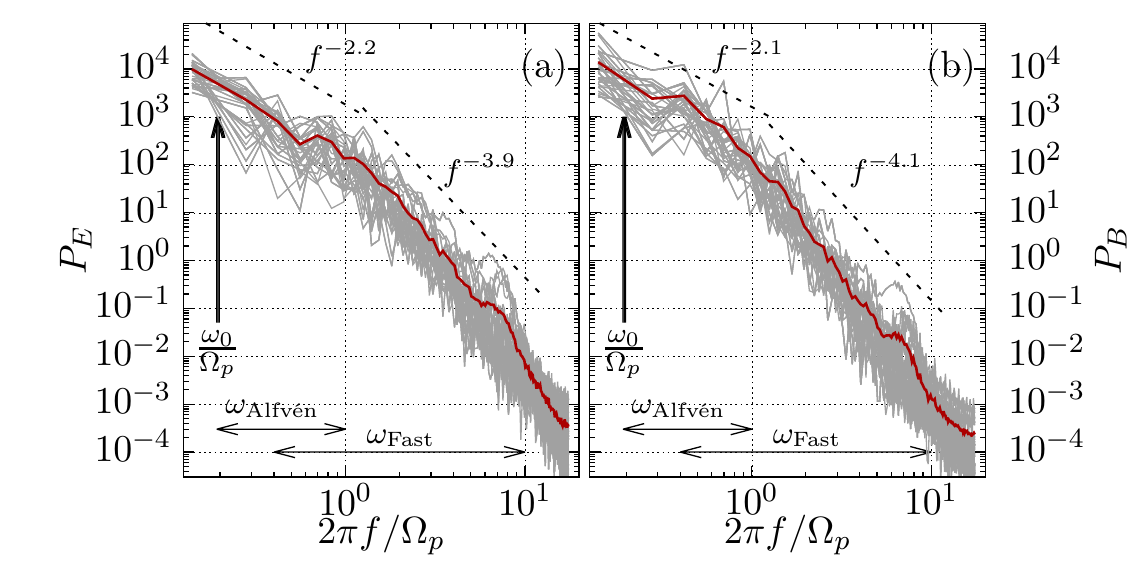}}%
  \caption{(a,b) Frequency power spectral density for electric and
    magnetic fields extracted from the 64 spatial points throughout
    the \texttt{HVM} simulation used in this work (grey). \change{The
      initialized frequencies $\omega_0/\Omega_p$ are indicated with
      an arrow on the left-hand side. The frequency ranges accessible
      to the Alfv\'en and fast dispersion surfaces for this
      simulation are indicated with horizontal arrows.}}
\label{fig:spectra}
\end{figure}

The simulation was evolved to $t_{max}=45\Omega_{p}^{-1}$. We selected
64 points, $\V{r}_0$, in the simulation's 3D spatial domain, producing
output of the electromagnetic fields $\V{E}'(\V{r}_0,t)$ and
$\V{B}'(\V{r}_0,t)$ in the simulation frame of reference as well as
the 3V proton velocity distribution $f_p(\V{r}_0,\V{v},t)$ at each of
the selected points. To demonstrate that there is significant power
distributed across a broadband range of frequencies, rather than being
composed of a handful of monochromatic \Alfven waves, we plot in
Fig.~\ref{fig:spectra} the frequency power spectral density for the
electric and magnetic field at each of the 64 spatial points. We see a
broad distribution of power across frequency at each point,
\change{rather than a peak at $2 \pi f_0/\Omega_p =
  \omega_0/\Omega_p$, where $\omega_0$ are the initialized Alfven
  frequencies, $\omega_0(k_\perp d_p =0.2,k_\parallel d_p = 0.2) =
  0.192 \Omega_p$, $\omega_0(k_\perp d_p =0.282,k_\parallel d_p = 0.2)
  = 0.195 \Omega_p$, and $\omega_0(k_\perp d_p =0.4,k_\parallel d_p =
  0.2) = 0.198 \Omega_p$.} \changetwo{Comparing this frequency
  distribution to the initial frequencies indicates
  significant nonlinear energy transfer from the initialized
  modes, producing a broadband turbulent system.}  \changetwo{The time series
  from which the frequency power spectra are calculated are stationary
  in the turbulent simulation, rather than traversing it at
  super-Alfv\'enic speeds as is typical of in situ measurements of the
  solar wind. As such, these single-point spectra do not capture the
  underlying spatial structure of the plasma fluctuations, which
  requires either invoking Taylor's Hypothesis, that the plasma-frame
  frequency is small compared to spatial advection
  \citep{Taylor:1938,Howes:2014a}, or measuring the system at multiple
  spatial points \citep{Klein:2019:WP}.}

\change{We compare the observed broadband distribution of frequencies
  to frequency ranges accessible to the Alfv\'en and fast normal mode
  solutions within the simulation's wavevector range $0.2 \le k_\perp
  d_p \le 3.2$ and $0.2 \le k_\parallel d_p \le 3.2$, calculated using
  the \textit{PLUME} dispersion solver, see
  Fig.~\ref{fig:freq_one}. Alfv\'en solutions are limited to a
  relatively narrow range of frequencies, $\omega/\Omega_p \in
  [.19,1.0]$. Above this frequency, we see a significant break in the
  power spectral densities in Fig.~\ref{fig:spectra}, indicating that
  there is relatively little power in higher frequency, non-Alfv\'enic
  fluctuations.}  \changetwo{Integrating the power in the electric and
  magnetic fluctuations in the Alfv\'en and fast frequency ranges, we
  find that nearly 95\% of the total power is contained at Alfv\'enic
  frequencies, with less than 30\% found in the partially overlapping
  fast frequency range.}
\change{Further discussion of wave mode
  identification using single point timeseries can be found in
  Appendix~\ref{app:modes}.}


\section{Applying the Field-Particle Correlation Technique}
\label{sec:fpem}

This section provides a brief overview of the field-particle
correlation technique. The field-particle correlation analysis
captures how energy is transferred between charged particles and
electromagnetic fields by correlating the structure of the particle
velocity distribution function with the electric field.  Applications
of this technique to simulations have been limited to velocity
distributions in one or two dimensions. Here we discuss the
application of the field-particle correlation technique to three
dimensional velocity distributions generated by the \texttt{HVM} code.

\subsection{Overview of Field-Particle Correlations}
\label{ssec:fpem}

For a collisionless magnetized plasma, the Vlasov equation
\begin{equation}
\frac{\partial f_s}{\partial t} + \mathbf{v}\cdot \nabla f_s + \frac
     {q_s}{m_s}\left[ \mathbf{E}+ \frac{\mathbf{v} \times \mathbf{B}
       }{c} \right] \cdot \frac{\partial f_s}{\partial \mathbf{v}} =
     0
\label{eqn:vlasov}
\end{equation}
describes the time evolution of the velocity distribution function of
charged particles of each species $s$, $f_s(\V{r},\V{v},t)$.  Combined
with Maxwell's equations, the Vlasov-Maxwell system describes the
self-consistent dynamics of a collisionless plasma.  We want to
measure the time rate of change of the microscopic kinetic particle
energy, $W_s(t) \equiv \int d\V{r} \int d\V{v} \ m_s v^2
f_s/2$. However, $\partial_t W_s$ can only be calculated by
integrating over all of 3D-3V phase space. Such a calculation is
accessible to numerical simulations, but not to measurements made from
a single point in coordinate space, as is typical for \emph{in situ}
measurements of heliospheric plasmas, such as the solar wind.

We therefore choose to track the energy density at a single point in 3D-3V
phase space, $\Theta_s(\V{r},\V{v},t) \equiv m_s v^2
f_s(\V{r},\V{v},t)/2$, and its time rate of change, which is found by
multiplying the Vlasov equation by $m_s v^2/2$ and not performing any
integration:
\begin{equation}
  \frac{\partial \Theta_s(\V{r},\V{v},t)}{\partial t} = - \frac{m_sv^2}{2}\V{v}\cdot \nabla
  f_s - q_s\frac{v^2}{2} \mathbf{E} \cdot \frac{\partial f_s}{\partial
    \mathbf{v}} - \frac{q_s}{c}\frac{v^2}{2} \left(\mathbf{v} \times
  \mathbf{B}\right) \cdot \frac{\partial f_s}{\partial \mathbf{v}}.
  \label{eqn:dws}
\end{equation}
Of the three terms on the right-hand side of Eqn.~\ref{eqn:dws}, it
can be shown \citep{Howes:2017a} that only the electric field term
will contribute to the net transfer of energy between the
electromagnetic fields and particles: the first term is zero for
periodic or infinitely distant boundary conditions and does not
exchange energy between the fields and the distribution; and the
magnetic field in the third term does no work on the distribution.

Integrating by parts the second term over velocity yields the species
current density dotted into the electric field $\V{j}_s\cdot\V{E}$,
representing the work done by $\V{E}$ on $f_s$ or vice-versa. By not
integrating this term, we resolve the velocity-space structure of
energy density transfer. As different mechanisms preferentially
energize particles with different characteristic velocities, resolving
the velocity-space structure of the energy density transfer allows
damping mechanisms to be differentiated using measurements from a
single point in coordinate space.

In an electromagnetic system, to determine the net contribution of the
parallel and perpendicular electric field to the energization of a
species $s$, we calculate the correlations
\begin{equation}
  C_{E_\parallel} (\V{r},\V{v},t,\tau)= C\left(- q_s\frac{v_\parallel^2}{2}
  \frac{\partial \delta f_s(\V{r},\V{v},t)}{\partial
    v_\parallel},E_\parallel(\V{r},t)\right)
   \label{eq:cepar}
\end{equation}
\begin{equation}
  C_{E_\perp}(\V{r},\V{v},t,\tau) = C\left(- q_s\frac{v_{\perp 1}^2}{2}
  \frac{\partial \delta f_s(\V{r},\V{v},t)}{\partial {v}_{\perp
      1}},{E}_{\perp 1}(\V{r},t)\right) + C\left(- q_s\frac{v_{\perp
      2}^2}{2} \frac{\partial \delta f_s(\V{r},\V{v},t)}{\partial
    {v}_{\perp 2}},{E}_{\perp 2}(\V{r},t)\right).
   \label{eq:ceperp}
\end{equation}
The unnormalized correlation of discretely sampled timeseries $A$ and
$B$ with uniform spacing $\Delta t$ at time $t_i$ is defined as
\begin{equation}
C(t_i,\tau=N \Delta t)\equiv \frac{1}{N}\sum_{j=i-N/2}^{i+N/2}A_jB_j,
\label{eqn:corr}
\end{equation}
with correlation interval of length $\tau = N \Delta t$.  Parallel and
perpendicular are defined with respect to the background magnetic
field $\V{B}_0$, with $\perp_1$ and $\perp_2$ denoting the orthogonal
components in the plane perpendicular to $\hat{\V{b}}=\V{B}_0/|B_0|$. The
$v^2$ component in the electric field term of Eqn~\ref{eqn:dws} is
replaced by the square of the component of the velocity $v_i$
associated with the component of the field with which the distribution
is being correlated $E_i$, as the net velocity integration is zero for
the other two components, $v_j$ and $v_k$.  By averaging over a time
interval $\tau$ longer than the characteristic timescale of the
dominant oscillations, rather than calculating the instantaneous rate
of change $C_{E_l}(t_i,\tau=0)=-q_s v_l^2 E_l \partial_{v_l} f_s/2$,
the contribution due to any oscillatory energy transfer, which does
not contribute to the net energization of the distribution, largely
cancels out.

The spatial energy density transfer rate at a single point $\V{r}_0$
associated with a single component of the electric field $E_l$ is
given by integrating over 3V velocity space,
\begin{equation}
\frac{\partial \bar{w}_{E_l}}{\partial t} (\V{r}_0,t_i,\tau) \equiv \int
d\V{v} C_{E_l} (\V{r}_0,\V{v},t_i,\tau)
  \label{eqn:partialw}
  \end{equation}
and the accumulated spatial energy density transferred through time
$t$ is
\begin{equation}
\Delta \bar{w}_{E_l} (\V{r}_0,t,\tau) = \int dt'
\frac{\partial \bar{w}_{E_l}(\V{r}_0,t',\tau)}{\partial t}.
\label{eqn:Deltaw}
\end{equation}
All energy density quantities are normalized to the average energy
density at that point in space over the simulated time interval $T$,
$w_0(\V{r}_0)=\left <\int d\V{v} m_p \V{v}^2 f_p(\V{r}_0,\V{v},t)/2
\right>_T$, e.g. $\partial_t \bar{w}_{E_l}(\V{r}_0,t,\tau)=\partial_t
w_{E_l}(\V{r}_0,t,\tau)/w_0(\V{r}_0)$.

\subsection{Field-Particle Correlation Implementation}
\label{ssec:implementation}

Here we describe how we calculate the velocity-resolved energy density
transfer rate using the simulated proton distribution
$f_p(\V{r}_0,\V{v},t)$ and the simulation-frame fields
$\V{B}'(\V{r}_0,t)$, and $\V{E}'(\V{r}_0,t)$ at a single spatial point
$\V{r}_0$, one of the 64 points $\V{r}_0$ probed in the turbulent
\texttt{HVM} simulation described in Sec.~\ref{sec:hvm}. As discussed
in \cite{Howes:2017a}, $\partial_t \bar{w}_{E_l}$ is the same for
correlations calculated using the velocity derivative of the full
distribution $\partial_{v_i} f_s$ or a perturbed distribution
$\partial_{v_i} \delta f_s$, where the perturbed velocity distribution
$\delta f_s = f_s - F_{0,s}$ is computed by subtracting a suitably
time-averaged mean velocity distribution, $F_{0,s}=\left <
f_s\right>_t$, as long as $F_{0,s}$ is an even function of velocity.
Here we calculate $F_{0,p}(\V{r}_0,\V{v})=\left <
f_p(\V{r}_0,\V{v},t)\right>_T$ averaged over duration of the
simulation $T$ and use the perturbed distribution $\delta
f_p(\V{r}_0,\V{v},t)$ for all of our correlation
calculations\footnote{We leave to a later work a discussion of the
  effects of different choices of mean velocity distributions
  $F_{0,s}$.}.

The vector velocity derivatives $\partial_{\V{v}} \delta
f_p(\V{r}_0,\V{v},t)$ are constructed using a centered-difference
method. The time-averaged bulk fluid velocity for a given point
$\V{U}(\V{r}_0)= \left <\V{v}_b(\V{r}_0,t) \right >_T$ is computed
using the instantaneous bulk velocity $\V{v}_b(\V{r}_0,t)
=[1/{n(\V{r}_0,t)}]$ $\int d\V{v} \ \V{v} f_p(\V{r}_0,\V{v},t)$ and the
instantaneous density $n(\V{r}_0,t)=\int d\V{v}f_p(\V{r}_0,\V{v},t)$.
Both $\partial_{\V{v}} \delta f_p(\V{r}_0,\V{v},t)$ and
$\V{E}'(\V{r}_0,t)$ are transformed to the frame of reference moving
at the average bulk flow velocity at each point, $\V{U}(\V{r}_0)$. For
the electric field, this requires applying the Lorentz transformation,
discussed for instance in \cite{Howes:2014a},
\begin{equation}
\V{E}=\V{E}'+\V{U}\times \V{B}/c,
\label{eqn:Eprime.hvm_fpc}
\end{equation}
where $\V{E}'$ is the electric field in the simulation frame, and
$\V{E}$ is the field in the average bulk flow frame.  Note that, under
the non-relativistic limit relevant to heliospheric plasmas, the
magnetic field requires no such transformation \citep{Howes:2014a},
i.e. $ \V{B}= \V{B}'$.

We define an instantaneous magnetic-field-aligned coordinate system at
position $\V{r}_0$ by parallel direction
$\hat{\V{b}}(\V{r}_0,t)=\V{B}(\V{r}_0,t)/|\V{B}(\V{r}_0,t)|$ and the
plane normal to $\hat{\V{b}}(\V{r}_0,t)$ spanned by in-plane unit
vectors $\hat{e}_{\perp 1}=\hat{\V{x}}\times \hat{\V{b}}(\V{r}_0,t)$
and $\hat{e}_{\perp 2}=\hat{\V{b}}(\V{r}_0,t)\times[\hat{\V{x}}\times
  \hat{\V{b}}(\V{r}_0,t)]$. We rotate the proton velocity distribution
$f_p$ and the electric field components into this field-aligned
coordinate systems. Note that, due to the large amplitude magnetic
field fluctuations required to achieve strong turbulence in this
Alfv\'en-ion cyclotron system, it is essential to project the fields
and particle velocities along the instantaneous magnetic field
direction to avoid smearing out of the resulting velocity-space
signatures of the energy transfer due to the variation in the magnetic
field direction over the correlation interval.

Using the electric field and proton velocity distribution in the
average bulk flow frame and field-aligned coordinates, we calculate
the parallel and perpendicular field-particle correlations using
Eqns.~\ref{eq:cepar} and \ref{eq:ceperp}, yielding the 3V
velocity-space resolved correlations $C_{E_\parallel}
(\V{r}_0,\V{v},t,\tau)$ and $C_{E_\perp} (\V{r}_0,\V{v},t,\tau)$.
We then integrate these correlations over 3V velocity space to
obtain the spatial energy density transfer rates, $\partial_t
\bar{w}_{E_\parallel} (\V{r}_0,t,\tau)$ and $\partial_t
\bar{w}_{E_\perp} (\V{r}_0,t,\tau)$, according to
Eqn.~\ref{eqn:partialw} and integrate those quantities over time to
obtain the accumulated spatial energy density changes, $\Delta
\bar{w}_{E_\parallel} (\V{r}_0,t,\tau)$ and $\Delta \bar{w}_{E_\perp}
(\V{r}_0,t,\tau)$, according to Eqn.~\ref{eqn:Deltaw}.

\begin{figure}
  \centerline{\includegraphics[width = 14cm]
    {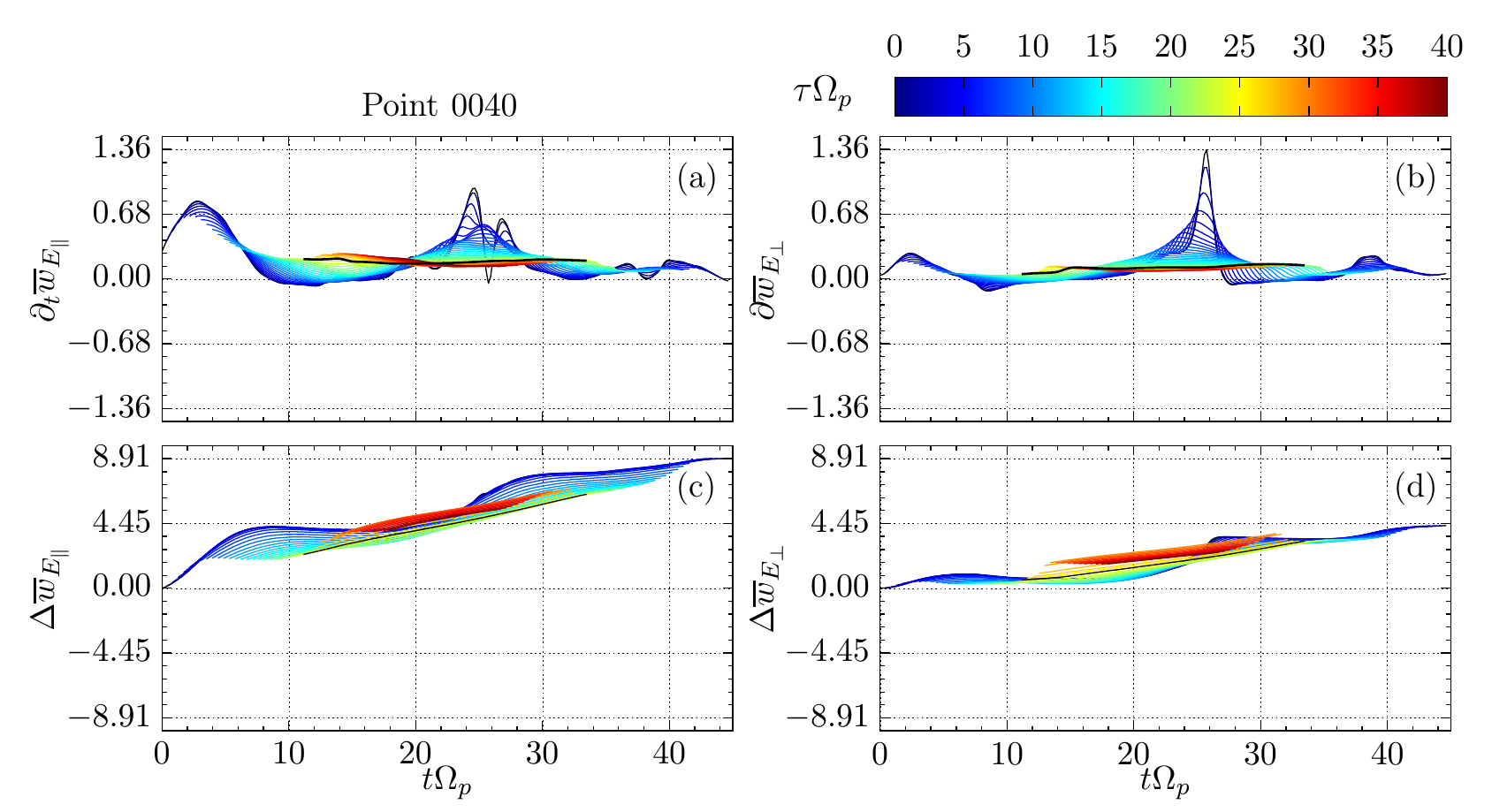}}%
  \caption{Velocity integrated correlations at a single point in the
    simulation domain for a range of correlation intervals
    $\tau\Omega_p$, indicated in color. (a,b) Energy density transfer
    rates $\partial_t \bar{w}_{E_\parallel}$ and $\partial_t
    \bar{w}_{E_\perp}$.  (c,d) Accumulated energy density transferred
    $\Delta \bar{w}_{E_\parallel}$ and $\Delta \bar{w}_{E_\perp}$. The
    thick black line indicates the correlation interval $\tau
    \Omega_p=22.5$ used in the remainder of this work.}
\label{fig:int}
\end{figure}

The next step is to determine a sufficiently long correlation time
interval $\tau$ over which to average in order to isolate the secular
component of the energy density transfer due to the electric field. In
this \texttt{HVM} turbulence simulation, the domain supports at the
largest scale MHD \Alfven waves that satisfy the dispersion relation
$\omega = k_\parallel v_A$.  In addition, as indicated by the \Alfven
mode wave phase velocities in Figure~\ref{fig:linear}(b) over the
range of resolved wavevectors (black box), the simulation also
supports higher frequency kinetic \Alfven waves at $k_\perp d_p>1 $
and lower frequency ion cyclotron waves at $k_\parallel d_p>1$.  In
previous studies \citep{Howes:2017a,Klein:2017b}, it was found that
averaging over intervals longer than the linear wave periods
associated with the transfer mechanisms of interest was sufficient to
isolate signatures of the secular transfer.  Note that the domain
scale MHD \Alfven waves initialized in the simulation have a frequency
$\omega = k_\parallel v_A= 2 \pi v_A/L_\parallel$, and therefore the
period of these waves, normalized to the proton cyclotron frequency,
is $T_0 \Omega_p = 2 \pi \Omega_p/\omega= L_\parallel \Omega_p/v_A =
10 \pi d_p \Omega_p/v_A \simeq 31.4$.  Here we substituted the domain
parallel length $ L_\parallel = 10 \pi d_p$ and have used the relation
between the proton inertial length and proton cyclotron frequency,
$d_p = v_A/\Omega_p$, to simplify the results.  With the period of
these largest-scale waves as guidance, we choose to test a range of
possible correlation intervals $ 0 \le \tau \Omega_p \le 40$.

In Figure~\ref{fig:int}, we plot $\partial_t \bar{w}_{E_l}$ and
$\Delta \bar{w}_{E_l}$ from a single spatial point over this range of
correlation intervals $ 0 \le \tau \Omega_p \le 40$. While the
instantaneous spatial energy density transfer rate ($\tau=0$, dark
blue) from $E_\parallel$ and $E_\perp$ varies significantly, we see
that as the correlation interval $\tau$ increases, this large
variation is reduced, leading to a smooth, net positive energy
transfer rate. \change{$\Delta \bar{w}_{E_l} (\V{r}_0,t,\tau)$ is
  adjusted to account for changes in the total integration time for
  varying correlation lengths, producing the expected convergent
  behavior.}

  To determine a sufficiently long
  interval $\tau$ to remove the oscillatory transfer we calculate the
  mean
and standard deviation
of $\partial_t \bar{w}_{E_\perp}$ and $\partial_t
\bar{w}_{E_\parallel}$ as a function of $\tau$ for all 64 spatial
points (not shown).  As expected by the form of the field-particle
correlation, the mean of the transfer rate is not significantly
affected by the choice of $\tau$, but the standard deviation is
reduced for longer correlation intervals. For a correlation interval
$\tau \Omega_p = 22.5$, the mean of the standard deviation, averaged
over the 64 output spatial points, of $\partial_t \bar{w}_{E_\perp}$
and $\partial_t \bar{w}_{E_\parallel}$ are reduced to less than $20
\%$ of the standard deviation for $\tau =0$. We therefore take the
interval $\tau \Omega_p=22.5$ to be the correlation length used
throughout this study; results are qualitatively similar to those
obtained using $\tau \Omega_p = 31.4$.

\section{Velocity-Space Signatures of Particle Energization}
\label{sec:vel}
In this section, we present the results of a field-particle
correlation analysis of proton energization occurring in the
Alfv\'en-ion cyclotron turbulence simulation described in
Sec.~\ref{sec:hvm}.  In particular, we present the first determination
of the typical velocity-space signature of proton cyclotron damping in
Sec.~\ref{sec:cyc}.  In addition, we analyze quantitatively the range
of variation of the velocity-space signatures of both proton cyclotron
damping and Landau damping in this simulation in Sec.~\ref{sec:quant}
and study the time variability in Sec.~\ref{ssec:distinguish}. This
section also demonstrates the key capability that the field-particle
correlation method can successfully employ single-point measurements
both to distinguish distinct mechanisms of energy transfer occurring
at the same point in space and to determine quantitatively the rates
of particle energization in each channel.

\subsection{Velocity-Space Signature of Cyclotron Damping}
\label{sec:cyc}
\begin{figure}
  \centerline{
    \includegraphics[width = 7cm, viewport = 50 25 170 120, clip=true]{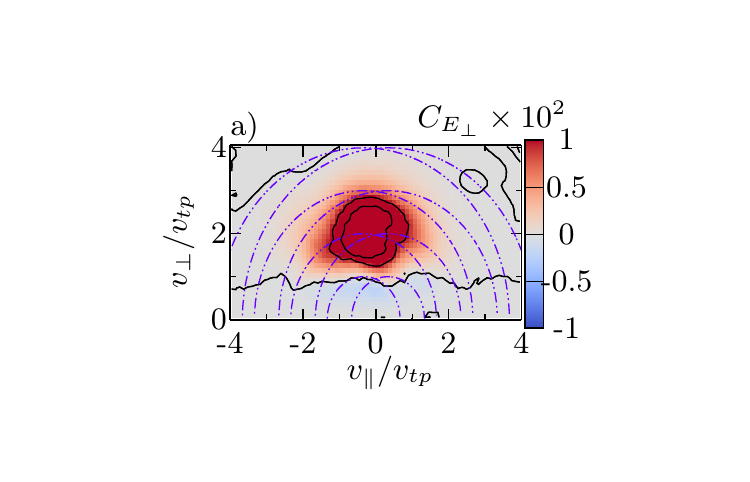}
    \includegraphics[width = 7cm, viewport = 50 25 170 120, clip=true]{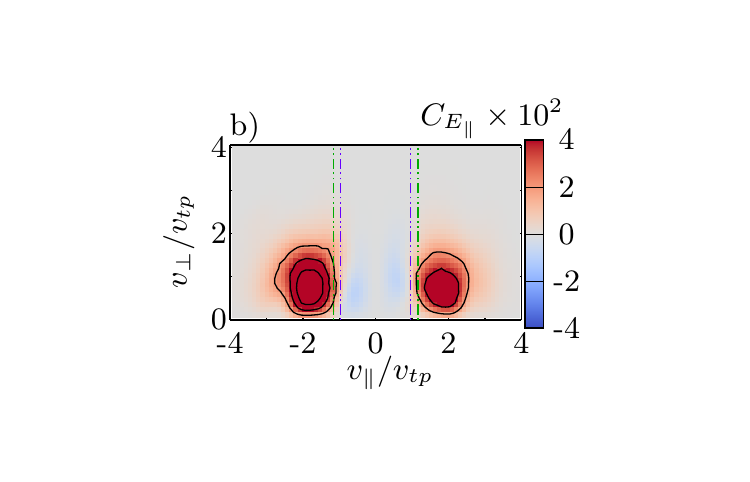}}
  \caption{Typical velocity-space signatures of (a) ion cyclotron
    damping and (b) ion Landau damping for point 40 at time
    $t\Omega_p=24.66$ using a correlation interval $\tau
    \Omega_p=22.5$, showing that the field-particle correlation
    technique can recover, using single-point measurements, the
    signatures of both energization mechanisms acting simultaneously
    at the same point in space. \change{Curved dashed lines in (a)
      indicate contours of constant energy in the ion-cyclotron
      wave-frame. The vertical dashed lines in (b) indicate the
      resonant velocities for the largest simulated scale (purple) and
      the most strongly Landau damped \Alfven waves (green).}}
\label{fig:vspacesig}
\end{figure}

Applying the perpendicular field-particle correlation $C_{E_\perp}$,
given by Eqn.~\ref{eq:ceperp}, to a single point in the Alfv\'en-ion
cyclotron turbulence simulation with a correlation interval $\tau
\Omega_p=22.5$, we plot the typical \emph{velocity-space signature of
  proton cyclotron damping}, shown in Fig.~\ref{fig:vspacesig}(a).
Here we have reduced the full 3V correlation
$C_{E_\perp}(v_{\parallel},v_{\perp,1},v_{\perp,2})$ to a 2V
correlation over gyrotropic velocity space by integrating over the
gyrophase angle $C_{E_\perp}(v_\parallel,v_\perp)=\int d \theta
v_\perp C_{E_\perp}(v_{\parallel},v_{\perp,1},v_{\perp,2})$ at time
$t\Omega_p=24.66$.  We find that protons are energized by the
perpendicular component of the electric field in a region of velocity
space with $1 \le v_\perp/v_{tp} \le 3$ and $-1.3 \le
v_\parallel/v_{tp} \le 1.3$ for the $\beta_p=1$ turbulence
simulation. This first demonstration of the velocity-space signature
of proton cyclotron damping in a kinetic simulation of plasma
turbulence is a key result of this study.

The location in velocity space of the cyclotron energization of the
protons generally agrees with predictions for the quasilinear
cyclotron diffusion plateau \citep{Kennel:1966,Marsch:2001,He:2015},
where the energy transfer mediated by $E_\perp$ is largest at the
confluence of the contours of constant energy for the forward and
backward propagating ion cyclotron waves, which satisfy
$\sqrt{(v_\parallel \pm \omega/k_\parallel)^2+v_\perp^2}=\mathcal{C}$.
In Fig.~\ref{fig:vspacesig}(a), we plot example contours (purple
dot-dashed) with $\mathcal{C}/v_{tp}=[1,2,3,4]$ for ion cyclotron
waves with $(k_\parallel d_p,k_\perp d_p)=(1,0.2)$, for which the
linear Vlasov-Maxwell dispersion relation yields a parallel phase
velocity $\omega/k_\parallel v_A=\omega/k_\parallel v_{tp}=0.335$ in
this $\beta_p=1$ plasma.

As shown in Fig.~\ref{fig:spectra}, this simulation generates a
broadband turbulent frequency spectrum.  The dispersive nature of the
Alfv\'en-ion cyclotron waves leads to a range of parallel phase
velocities (and thus a range of frequencies) $0.13 \le
\omega/k_\parallel v_A \le 0.96$ over the range of parallel
wavevectors in this simulation, $0.2 \le k_\parallel \rho_p \le 3.2$.
The centers of the sets of circular contours in
Fig.~\ref{fig:vspacesig}(a) would shift with this variation in
parallel phase velocities $\omega/k_\parallel$, potentially leading to
a smearing of the observed velocity-space signature.  Therefore, the
particular contours (purple) plotted in Fig.~\ref{fig:vspacesig}(a)
for ion cyclotron waves with $(k_\parallel d_p,k_\perp d_p)=(1,0.2)$
are merely presented as useful guide for the qualitative
interpretation of the velocity-space signature.

We also plot in Fig.~\ref{fig:vspacesig}(b) the parallel
field-particle correlation over 2V gyrotropic velocity space,
$C_{E_\parallel}(v_\parallel,v_\perp)$, given by Eqn.~\ref{eq:cepar},
for the same spatial point and using the same correlation interval
$\tau \Omega_p=22.5$ centered at the same time $t\Omega_p=24.66$.
Here we find that protons are energized by the parallel component of
the electric field in two regions of velocity space, defined by $ 1\le
|v_\parallel/v_{tp}| \le 2.5$ and $0 \le v_\perp/v_{tp} \le 1.5$.

To interpret quantitatively the location of the parallel energization
in velocity space, we plot vertical lines at the resonant parallel
phase velocity $\omega/k_\parallel v_{tp}$ for the domain-scale
\Alfven waves with $k_\perp d_p=0.2$ (purple) and for the kinetic
\Alfven waves with the peak proton Landau damping rate at $k_\perp
d_p=1.2$(green).  We find that the proton energization is negative
(blue) for parallel velocities less than the resonant phase velocities
$|v_\parallel/v_{tp}| < \omega/k_\parallel v_{tp}$ and is positive
(red) for parallel velocities greater than the resonant phase
velocities $|v_\parallel/v_{tp}| > \omega/k_\parallel v_{tp}$. This
typical bipolar signature of the energy transfer about the resonant
parallel phase velocity indicates that this collisionless energy
transfer is associated with the Landau resonance, consistent with
previous determinations of the velocity-space signature of the Landau
damping of \Alfven waves in single wave simulations
\citep{Howes:2017b,Klein:2017b}, gyrokinetic turbulence simulations
\citep{Klein:2017b}, and observations of the Earth's turbulent
magnetosheath \citep{Chen:2019}.  The \texttt{HVM} results here
represent an independent confirmation of the velocity-space signature
of Landau damping in Alfv\'en-ion cyclotron turbulence.

\begin{figure}
  \centerline{\includegraphics[width = 12cm]
    {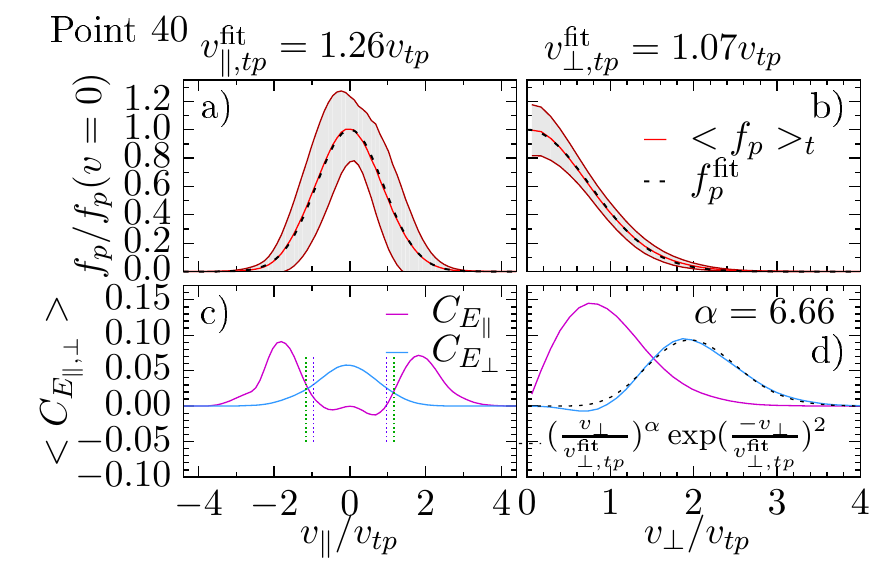}}%
  \caption{\change{Time-averaged reduced proton velocity distributions and
    their standard deviation measured at a single spatial point (a,b)
    and the associated reduced correlations $C_{E_{\perp,\parallel}}
    (v_\parallel)$ (c) and $C_{E_{\perp,\parallel}} (v_\perp)$
    (d). The vertical dashed lines in (c) indicate the dominant
    parallel resonant velocities for the simulation, while the black
    dashed line in (d) represents the best fit to $C_{E_{\perp}}
    (v_\perp)$.}}
\label{fig:1d_reduce}
\end{figure}

\change{ We further reduce the 2D gyrotropic velocity space to a
  function of either $v_\perp$ or $v_\parallel$ in
  Fig.~\ref{fig:1d_reduce}.  In this reduced space, we plot the proton
  distribution function measured at point 40 averaged over the
  duration of the simulation, as well as the standard deviation around
  the average value.  The structures of $C_{E_\parallel}(v_\parallel)$
  and $C_{E_\perp}(v_\perp)$ have the same shape as inferred from the
  gyrotropic representation.  By reducing the correlations to a function of
  $v_\perp$, we can compare the perpendicular heating to quasilinear
  predictions. If the perpendicular velocity diffusion coefficient
  associated with cyclotron heating is independent of perpendicular
  velocity, as predicted by \cite{Kennel:1966} and
  \cite{Isenberg:2007}, we would expect $C_{E_\perp} (v_\perp) \propto
  v_\perp^3 \exp(-v_\perp^2/v_{th}^2)$.  To test this prediction, we
  fit the average perpendicular thermal width of the reduced proton
  velocity distribution, $v^{\textrm{fit}}_{\perp,tp}$ and then fit
  $C_{E_\perp}(v_\perp)$ to the functional form
  $(v_\perp/v^{\textrm{fit}}_{\perp,tp})^\alpha
  \exp(-v_\perp^2/v^{\textrm{fit}}_{\perp,tp})^2$. We are able to
  extract a good fit from this procedure but find $\alpha \approx
  6.6$, rather than the expected value of $3$, qualitatively similar
  to the results presented in \cite{Arzamasskiy:2019}, indicating a
  strong dependence of the energy diffusion on $v_\perp$.  }
  
It is worth noting that the bipolar aspect of the energy transfer via
the Landau resonance is less apparent in this \texttt{HVM} simulation
of Alfv\'en-ion cyclotron turbulence than in previous analyses of
gyrokinetic simulations and magnetosheath observations.  This smearing
out of the velocity-space signature may be due to the perpendicular
motions of the large-amplitude \Alfven waves with $\delta B_\perp \sim
B_0$ in the \texttt{HVM} simulation. \change{These relatively large
  amplitude Alfv\'enic fluctuations lead to significant shifts in the
  proton velocity distribution from the average bulk velocity frame,
  as seen in the width of the standard deviation about the
  time-averaged VDF in Fig.~\ref{fig:1d_reduce}(a,b), broadening the
velocity-space regions over which energy is transferred.} For the
anisotropic \Alfvenic fluctuations with $k_\parallel \ll k_\perp$ in
gyrokinetic simulations and in the dissipation-range turbulence of the
magnetosheath, strong turbulence can be achieved with $\delta B_\perp
\ll B_0$, possibly leading to a more clear bipolar velocity-space
signature, because the smaller amplitude of the turbulent fluctuations
would lead to less smearing of the characteristic bipolar appearance.

The velocity-space signatures of (a) cyclotron damping and (b) Landau
damping, computed using single-point measurements of the electric
field and proton velocity distribution over the same correlation time
interval and at the same position in space, clearly demonstrate a
second key result of this study: that the field-particle correlation
method can successfully employ single-point measurements to
distinguish distinct mechanisms of energy transfer occurring at the
same point in space.

Of course, since the parallel correlation $C_{E_\parallel}$ integrated
over velocity simply yields $j_{\parallel p } E_\parallel$, and the
velocity-integrated perpendicular correlation $C_{E_\perp}$ yields
$\V{j}_{\perp p } \cdot \V{E}_\perp$, one could argue that this
separation of cyclotron from Landau energization mechanisms could
simply be achieved by separating the parallel and perpendicular
components of $\V{j} \cdot \V{E}$.  However, determining the
components of $\V{j} \cdot \V{E}$ provides only the rate of change of
spatial energy density due to the different components of $\V{E}$, but
nothing about the specific physical mechanism responsible for this
energy transfer. The field-particle correlations $C_{E_\parallel}
(\V{v},t,\tau)$ and $C_{E_\perp}(\V{v},t,\tau)$, because
they provide the variation of the energization as a function of
particle velocity, yield vastly greater detail about the mechanisms
through their velocity-space signatures, with the possibility to
distinguish one mechanism from another through qualitative or
quantitative differences in the characteristic velocity-space
signatures of each mechanism.

For example, proton cyclotron damping in a $\beta_p=1$ plasma, as
shown in Figs.~\ref{fig:vspacesig}(a) and ~\ref{fig:1d_reduce}(c,d),
is expected to energize protons with velocities $1 \le v_\perp/v_{tp}
\le 3$ and $-1.3 \le v_\parallel/v_{tp} \le 1.3$.  Landau damping in a
$\beta_p=1$ plasma, on the other hand, is expected to energize protons
with velocities $1\le |v_\parallel/v_{tp}| \le 2.5$ and $0 \le
v_\perp/v_{tp} \le 1.5$, with a bipolar signature changing sign about
the parallel resonant phase velocity. These detailed quantitative
features enable one to identify the specific physical mechanisms
responsible for the energization.  Ongoing work to determine the
velocity-space signatures of different energization mechanisms,
including their variation as a function of the plasma parameters such
as $\beta_p$, will provide a framework for the interpretation of the
velocity-space signatures obtained through the field-particle
correlation analysis of both kinetic numerical simulations and
spacecraft observations, potentially providing a clear procedure for
the identification of the particle energization mechanisms that play a
role in the dissipation of turbulence in these systems.

\subsection{Variation of Velocity-Space Signatures}
\label{sec:quant}

Now that we have presented fiducial velocity-space signatures for
cyclotron damping and Landau damping in Fig.~\ref{fig:vspacesig}, we
seek to quantify the variation of the velocity-space signatures of the
2V gyrotropic perpendicular and parallel correlations
$C_{E_\perp}(v_\parallel,v_\perp)$ and
$C_{E_\parallel}(v_\parallel,v_\perp)$ in our \texttt{HVM} simulation of
Alfv\'en-ion cyclotron turbulence.

Intuition gained from plane-wave studies of linear collisionless
damping of waves often leads people to believe that linear
collisionless damping is expected to occur uniformly in space.  This
belief is not correct. Similar to the case that any spatially varying
waveform can be decomposed into its plane-wave components, the spatial
distribution of energy transfer associated with linear collisionless
damping mechanisms is controlled by the spatial distribution of the
field doing the work, and this field may arise in a spatially
non-uniform manner if numerous plane-wave modes contribute to the
waveform of the field. In plasma turbulence, early studies discovered
that intermittent current sheets naturally develop
\citep{Matthaeus:1980,Meneguzzi:1981}, and more recent work has shown
that the dissipation of turbulent energy is largely concentrated near
these current sheets
\citep{Uritsky:2010,Osman:2011,Zhdankin:2013,Navarro:2016}.  Although
the idea of dissipation in current sheets suggests a possible role of
magnetic reconnection, in fact, a recent study has shown a clear
counterexample in which collisionless wave-particle interactions
underlie spatially non-uniform energy transfer. In a gyrokinetic
simulation where strongly nonlinear \Alfven wave collisions
\citep{Howes:2013a} self-consistently generate current sheets
\citep{Howes:2016b}, spatially non-uniform particle energization
occurs, with greater energy transfer near current sheets, but the
underlying mechanism of energy transfer in this case is clearly
identified, using the field-particle correlation technique, as Landau
damping \citep{Howes:2018a}.  Therefore, even if the removal of energy
from turbulent fluctuations occurs dominantly through collisionless
wave-particle interactions, one would expect that the net energy
transfer would vary significantly from point to point in a strongly
turbulent system.

Furthermore, collisionless energy transfer via wave-particle
interactions is reversible, meaning that in addition to positive
energy transfer from the fields to the particles, one can also find
regions of negative energy transfer from the particles to the fields.
Nonetheless, the regions of negative energy transfer mediated by
collisionless wave-particle interactions still yield velocity-space
signatures characteristic of the energy transfer mechanism, but with
opposite sign \citep{Howes:2018a}.  Here we hope to explore the
typical variation in space of the velocity-space signatures of the
perpendicular and parallel field-particle correlations, examining
regions of positive energy transfer, negative energy transfer, and
negligible energy transfer.

\begin{figure}
  \centerline{\includegraphics[width = 14cm, viewport = 50 20 310 275,
      clip=true]
    {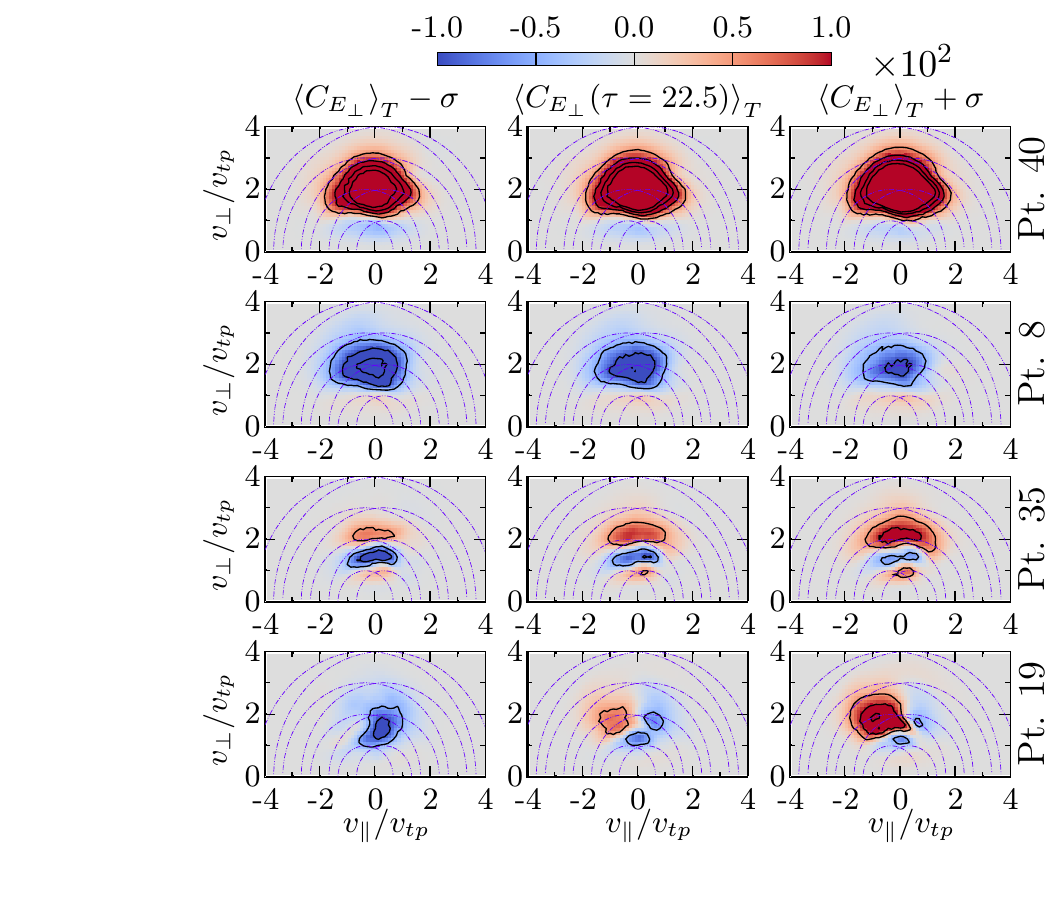}}
  \caption{Mean secular energy-density transfer rate between the
    protons and $E_\perp$ at four points throughout the \texttt{HVM}
    simulation, $\left <C_{E_\perp}\right>_T$, center column. The velocity-dependent
    standard deviation at each point is added to or subtracted from
    the mean in the right and left columns. Contours of constant
    energy in the wave frame of forward and backward propagating
    Alfv\'en-Ion cyclotron waves are shown in purple.}
\label{fig:CD_gyro}
\end{figure}

Here we characterize the variations of the perpendicular and parallel
correlations $C_{E_\perp}(v_\perp,v_\parallel)$ and
$C_{E_\parallel}(v_\perp,v_\parallel)$ at four different spatial
points in our \texttt{HVM} turbulence simulation, representing cases
with significant energy transfer either direction between the protons
and $E_\perp$ or $E_\parallel$, as well as cases with relatively little
net energy transfer. To quantify the variation in the velocity-space
signature, we use a correlation interval $\tau \Omega_p=22.5$ to
compute the correlation $C_{E_\perp}(v_\perp,v_\parallel,t,\tau)$ at
each point as a function of the time $t$ at the center of the
correlation interval.  We compute the mean of this correlation over
the entire simulation time $T$, $\langle C_{E_\perp} \rangle_T$, and
the standard deviation of its variation $\sigma(C_{E_\perp})$ at each
point in gyrotropic velocity space $(v_\perp,v_\parallel)$. To
visualize the variation in time at each of the four points, we plot in
Fig.~\ref{fig:CD_gyro} the mean value in the central column, the mean
minus the standard deviation at each point (left column), and the mean
plus the standard deviation (right column).

Regardless of the sign or amplitude of $C_{E_\perp}$, we see in
Fig.~\ref{fig:CD_gyro} that the transfer associated with $E_\perp$ is
strongly concentrated between $1 \le v_\perp/v_{tp} \le 3$ and $-2 \le
v_\parallel/v_{tp} \le 2$. We plot the same contours (purple) of
constant $\sqrt{(v_\parallel \pm
  \omega/k_\parallel)^2+v_\perp^2}=\mathcal{C}$ for ion cyclotron
waves used in Fig.~\ref{fig:vspacesig} with $(k_\parallel d_p,k_\perp
d_p)=(1,0.2)$ as a guide for interpretation.
  
At point 40 in Fig.~\ref{fig:CD_gyro}, where we find significant
energy transfer to the protons, we observe that protons with $v_\perp
> v_{tp}$ gain a significant amount of energy while protons with
$v_\perp < v_{tp}$ lose a relatively small amount of energy to
$E_\perp$.  At point 8, where we observe energy transfer from the
protons, the pattern remains similar, but with the signs reversed.  At
points 35 and 11 where the net spatial energy density transfer
(integrated over all velocity space) is relatively small, we see two
distinct behaviors.  At point 35, we find regions of strong energy
density transfer of opposite sign in adjacent bands of $v_\perp$ that
approximately follow contours of constant energy. When integrated
over velocity, the opposite signs of these bands significantly
reduce the net transfer. At point 11, the sign of $C_{E_\perp}$
changes part-way through the simulation, leading to little net energy
transfer when averaged over the full simulation time.

\begin{figure}
  \centerline{\includegraphics[width = 14cm, viewport = 50 20 310 275,
      clip=true]
    {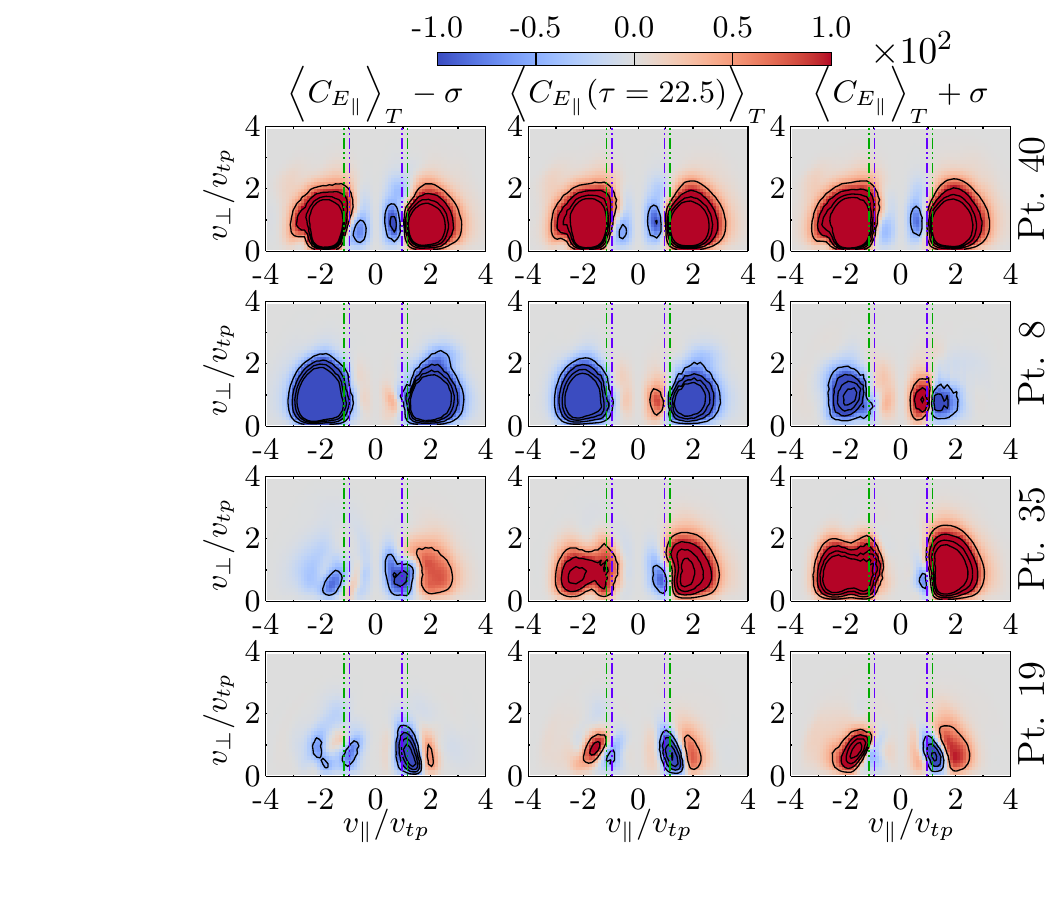}}
  \caption{Mean secular energy-density transfer rate between the
    protons and $E_\parallel$ at the same four points considered in
    Fig.~\ref{fig:CD_gyro}, $\left<C_{E_\parallel}\right>_T$, center column, with the
    velocity-dependent standard deviation added to or subtracted from
    the mean in the right and left columns. The Landau resonant
    velocity for the largest simulated and the most strongly Landau
    damped \Alfven waves are shown in purple and green.}
\label{fig:LD_gyro}
\end{figure}

In Fig.~\ref{fig:LD_gyro}, we present the quantitative analysis of the
variation of the parallel correlation
$C_{E_\parallel}(v_\perp,v_\parallel)$ at the same four points
considered in Fig.~\ref{fig:CD_gyro}, with the mean over the entire
simulation time $\langle C_{E_\parallel} \rangle_T$ in the center
column, and minus or plus the standard deviation in the left and right
columns, respectively. Here we find that the energy transfer is
concentrated in two regions, defined by $0.5 \le|v_\parallel/v_{tp}|
\le 3$ and $0 \le v_\perp/v_{tp} \le 2$.

At all four points, we see some evidence of the bipolar resonant
signature associated with Landau damping seen in previous numerical
\citep{Howes:2017b,Klein:2017b,Howes:2018a} and observational studies
\citep{Chen:2019}. The change of sign is generally consistent with the
resonant parallel phase velocities of the largest-scale \Alfven waves
in the system (purple) and the most strongly damped kinetic \Alfven
waves with $k_\perp d_p=1.2$ (green). At points 40, 35, and 11, net
energy is transferred from $E_\parallel$ to the protons, with positive
energy transfer at parallel velocities above the resonant velocity and
negative energy transfer below.  At point 8, where there is net
transfer from the protons to $E_\parallel$, the bipolar pattern of
resonant collisionless energy transfer is the same, but the signs of
the energy transfer are reversed, consistent with a previous study of
Landau-resonant energization in current sheets generated by strong
\Alfven wave collisions \cite{Howes:2018a}.

The careful reader will note a difference in the widths of the regions
of resonant transfer between this simulation and the
strong gyrokinetic turbulence simulation described in
\cite{Klein:2017b}. As $\delta B_\perp/B_0$ is necessarily much larger
for this wavevector-isotropic system in order for the simulation to
satisfy $\chi = (k_\perp/k_\parallel) (\delta B_\perp/ B_0) \approx
1$, the proton distribution is more perturbed, resulting
in a broadening of the resonant signature. Studies of the
effect of variations in $k_\perp/k_\parallel$ and $\delta B_\perp/
B_0$ will be left to future work.

In summary, we find that although the amplitude and sign of the
energization of particles by $E_\perp$ and $E_\parallel$ varies from
position to position in strong turbulence, the regions of velocity
space where particles participate in the energy transfer remain
remarkably constant.  Furthermore, the velocities at which the energy
transfer changes sign also appear to be reproducible from point to
point.  This pattern of energy transfer in velocity-space, denoted the
\emph{velocity-space signature}, provides a valuable tool for the
identification of the physical mechanisms responsible for the removal
of energy from turbulent fluctuations and consequent energization of
particles. Further work is needed to determine how these
velocity-space signatures change quantitatively as a function of the
plasma parameters, in particular the plasma $\beta_p$, which controls
where the wave phase velocities fall within the thermal distribution
of particle velocities.

It is worthwhile to note that the different regions in
velocity space of energy transfer for $E_\parallel$ and $E_\perp$ are
partly enforced by the mathematical form of the correlations,
Eqns.~\ref{eq:cepar} and \ref{eq:ceperp}. For example, the presence of
the $v_\parallel^2$ term in $C_{E_\parallel}$ dictates that the
parallel energy transfer must drop to zero as $|v_\parallel|
\rightarrow 0$, and similarly the presence of the $v_{\perp 1}^2$ and
$v_{\perp 2}^2$ factors in $C_{E_\perp}$ require that the
perpendicular energy transfer drops to zero as $|v_\perp| \rightarrow
0$. Nonetheless, the mathematical forms of $C_{E_\parallel}$ and
$C_{E_\perp}$ are simply the terms for the parallel and perpendicular
energy transfer in the equation for the evolution of the phase-space
energy density, Eqn.~\ref{eqn:dws}. Therefore, the results of the
correlation can be interpreted directly in physical terms, where the
unnormalized correlation is precisely the rate of change of
phase-space energy density at each point in 3D-3V phase space.  A key
additional point to emphasize is that the change of sign of the energy
transfer---such as that frequently found at the parallel resonant
velocity in $C_{E_\parallel}(v_\parallel,v_\perp)$---is \emph{not}
guaranteed by the mathematical form of the correlation. This feature
is therefore indicative of the governing physical mechanism,
suggesting that such features in
the velocity-space signatures of different mechanisms can be used to
identify the mechanisms dominating particle energization in both
numerical simulations and single-point spacecraft observations.

\subsection{Time Evolution of $C_{E_\perp}$ and $C_{E_\parallel}$}
\label{ssec:distinguish}

The 2V gyrotropic velocity-space signatures presented in
Figs.~\ref{fig:vspacesig}--\ref{fig:LD_gyro} provide valuable
information about the energy transfer as a function of particle
velocity but do not contain information about the variation of the
energy transfer as a function of time.  Timestack plots of the reduced
perpendicular correlation $C_{E_\perp}(v_\perp,t) = \int d v_\perp
C_{E_\perp}(v_\parallel,v_\perp,t)$ and reduced parallel correlation
$C_{E_\parallel}(v_\parallel,t) = \int d v_\parallel
C_{E_\parallel}(v_\parallel,v_\perp,t)$ enable the energy transfer to
be visualized as a function of time and the most relevant component of
velocity space. The motivation of these particular reductions is the
strong dependence of $C_{E_\perp}$ on $v_\perp$ and weak dependence on
$v_\parallel$, as seen in Fig~\ref{fig:CD_gyro}; similarly,
$C_{E_\parallel}$ has a strong dependence on $v_\parallel$ and a weak
dependence on $v_\perp$, as seen in Fig.~\ref{fig:LD_gyro}.  We have
therefore not included plots of $C_{E_\perp}(v_\parallel)$ and
$C_{E_\parallel}(v_\perp)$.

\begin{figure}
  \centerline{\includegraphics[width = 14cm, viewport = 25 5 300 165,
      clip=true]
    {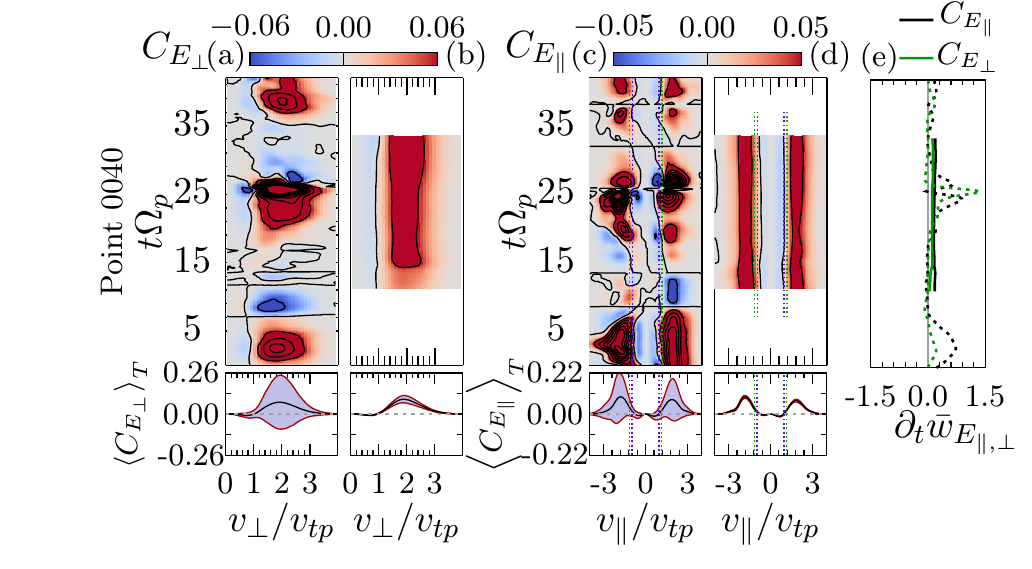}}
  \caption{Reduced field-particle correlations $C_{E_\perp}(v_\perp)$
    with (a) $\tau = 0$ and (b) $\tau \Omega_p = 22.5$ and
    $C_{E_\parallel}(v_\parallel)$ with (c) $\tau = 0$ and (d) $\tau
    \Omega_p = 22.5$. The lower panels of columns (a)-(d) show the
    time-averaged, velocity-dependent energy transfer rate, with the
    mean in black and one standard deviation in red. (e) The
    velocity-integrated spatial energy density transfer rates
    $\partial_t \bar{w}_{E_\perp}$ (green) and $\partial_t
    \bar{w}_{E_\parallel}$ (black) are shown for $\tau \Omega_p=0$
    (dashed) and $22.5$ (solid).}
\label{fig:reduced.A}
\end{figure}

\begin{figure}
  \centerline{\includegraphics[width = 14cm, viewport = 25 5 300 165,
      clip=true]
    {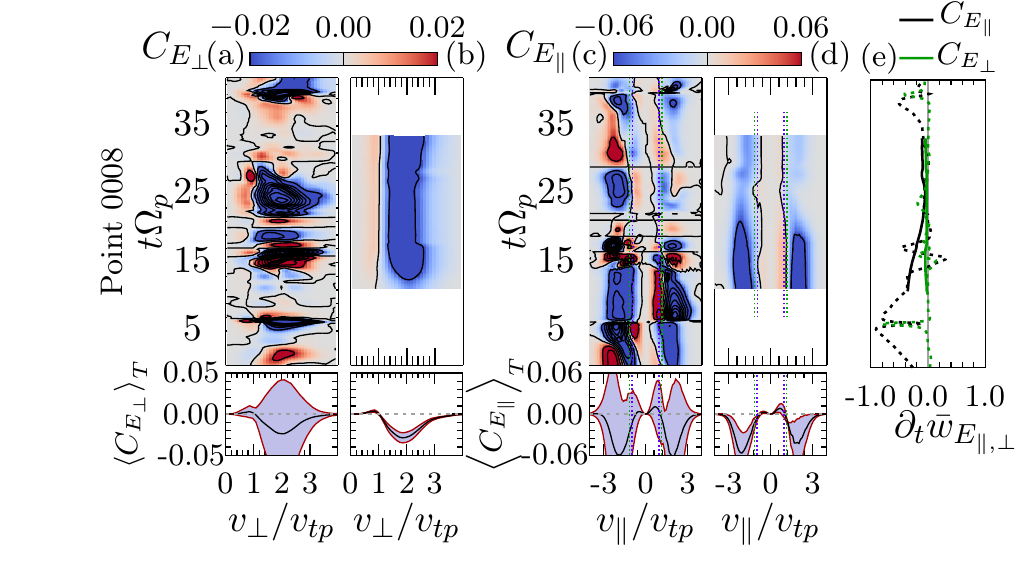}}
  \caption{Reduced field-particle correlations, organized in the same
    format at Figure~\ref{fig:reduced.A} but for point 8.  At this
    point, there is a net transfer of energy from the protons to
    $E_\parallel$ and $E_\perp$, with $E_\parallel$ receiving more energy.}
\label{fig:reduced.B}
\end{figure}

\begin{figure}
  \centerline{\includegraphics[width = 14cm, viewport = 25 5 300 165,
      clip=true]
  {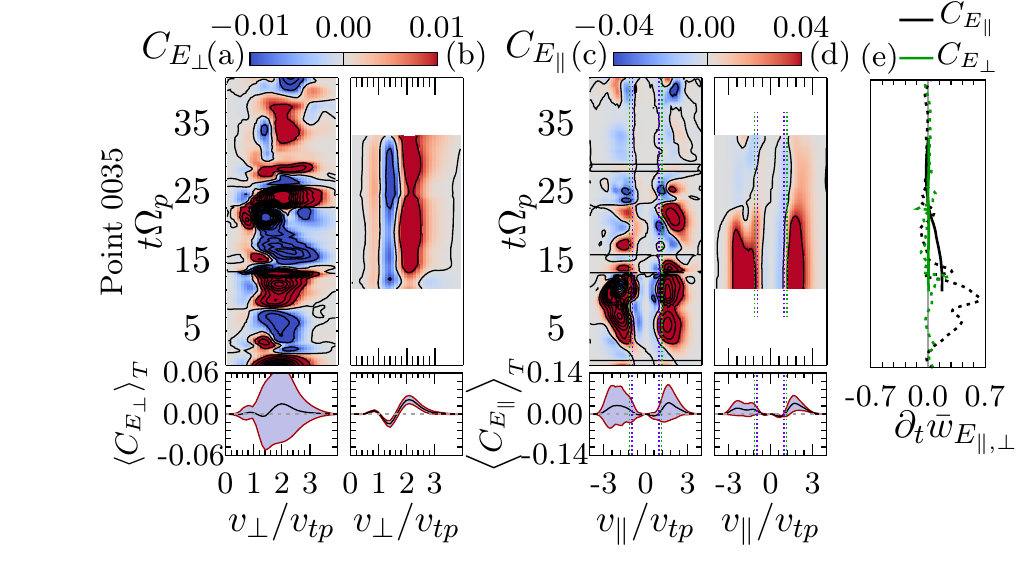}}
  \caption{Reduced field-particle correlations, organized in the same
    format at Figure~\ref{fig:reduced.A} but for point 35.  At this
    point, there is a net transfer of energy to the protons via
    $E_\parallel$, and very little transfer of energy mediated by
    $E_\perp$.}
\label{fig:reduced.C}
\end{figure}

\begin{figure}
  \centerline{\includegraphics[width = 14cm, viewport = 25 5 300 165,
      clip=true]
    {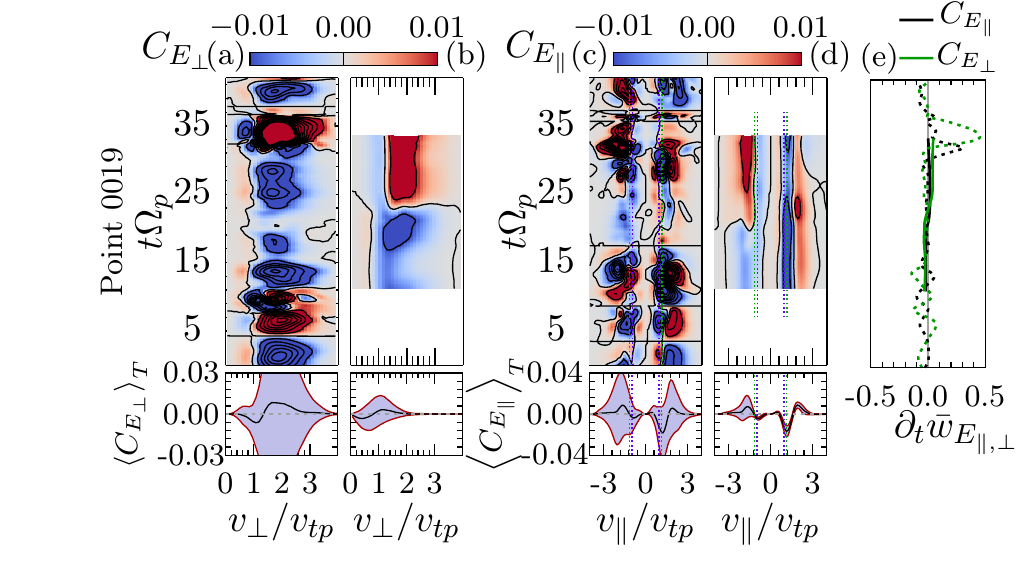}}
  \caption{Reduced field-particle correlations, organized in the same
    format at Figure~\ref{fig:reduced.A} but for \change{point 19}.  At this
    point, there is relatively little net transfer of energy between
    the protons and the electric field.}
\label{fig:reduced.D}
\end{figure}

In Figs~\ref{fig:reduced.A} through ~\ref{fig:reduced.D}, we consider
the same four spatial points highlighted earlier in
Figs~\ref{fig:CD_gyro} and \ref{fig:LD_gyro}. In each figure, columns
(a) and (b) present timestack plots of $C_{E_\perp}(v_\perp,t;\tau=0)$
and $C_{E_\perp}(v_\perp,t;\tau\Omega_p=22.5)$. Plotted at the bottom
of each column is the mean value averaged over the entire simulation
time $T$, $\left< C_{E_\perp}(v_\perp, t,\tau)\right>_T$ (black), with
the extent of the standard deviation about the mean (shaded). Columns
(c) and (d) present the same for
$C_{E_\parallel}(v_\parallel,t;\tau=0)$ and
$C_{E_\parallel}(v_\parallel,t;\tau\Omega_p=22.5)$. Column (e)
presents velocity-integrated energy density transfer rates,
$\partial_t \bar{w}_{E_\parallel}$ (black) and $\partial_t
\bar{w}_{E_\perp}$ (green) for $\tau=0$ (dashed) and
$\tau\Omega_p=22.5$ (solid).  Note that $\partial_t
\bar{w}_{E_\parallel}$ is equal to the work done by the parallel
electric field on the protons, $j_{\parallel,p} E_\parallel$, and
$\partial_t\bar{w}_{E_\perp}$ is equal to the work done by the
perpendicular electric field on the protons, $\V{j}_{\perp,p} \cdot
\V{E}_\perp$.


Several key points can be gleaned from the set of timestack plots in
Figs~\ref{fig:reduced.A}--\ref{fig:reduced.D}.  First, in all cases,
throughout the evolution of the simulation, the perpendicular energy
transfer diagnosed by $C_{E_\perp} (v_\perp)$ falls primarily in the
range $1 \le v_\perp/v_{tp} \le 3$ and the parallel energy transfer
diagnosed by $C_{E_\parallel} (v_\parallel)$ falls primarily within
the two ranges $ -3 \le v_\parallel/v_{tp} \le -0.5$ and $0.5 \le
|v_\parallel/v_{tp}| \le 3$, consistent with the findings of the 2V
gyrotropic velocity space signatures in Sec.~\ref{sec:quant}. Second,
as shown in the lower panels of columns (a) and (c) in the figures,
the instantaneous energy transfer rate (the correlation with $\tau=0$)
experiences a wide variation (shaded region) of $C_{E_\perp}
(v_\perp)$ and $C_{E_\parallel} (v_\parallel)$ in time, consistent
with the idea of a significant \emph{oscillating energy transfer}
\citep{Howes:2017a}.  This oscillating energy transfer largely
averages out over sufficiently long correlation intervals, as shown in
the lower panels of columns (b) and (d) for $\tau \Omega_p=22.5$,
where the variation of the energy transfer rate is greatly diminished,
as intended with the field-particle correlation method. This removal
of the oscillating component can be seen clearly in the
velocity-integrated spatial energy transfer rates $\partial_t
\bar{w}_{E_\parallel}$ and $\partial_t \bar{w}_{E_\perp}$ in column
(e), where the amplitude of the energy transfer in the instantaneous
case ($\tau=0$, dashed) is greatly reduced when a sufficiently long
correlation interval is chosen ($\tau \Omega_p=22.5$, solid).

Third, for the two points, 35 and 11, at which there is little net
energy transfer by the perpendicular electric field, we find two
different behaviors. Although, as shown in panel (a) both cases
display significant instantaneous transfers of energy at various
points in velocity and time, the cancellation of these positive and
negative transfers is different in the two cases: (i) at point 35, the
energy transfer varies as a function of $v_\perp$, so that the
velocity-integrated energy transfer remains small at all times; and
(ii) at point 11, the velocity-integrated energy transfer is positive
at early times and negative at late times, so that, when averaged over
time (lower panel, column (a)), the net energy transfer is small.

Finally, considering all 64 spatial points diagnosed in the simulation
(not shown), the fraction of the energy density transfer mediated by
$E_\parallel$ compared to the total transfer rate, $|\partial_t
\bar{w}_{E_\parallel}|/|\partial_t \bar{w}_{E_\parallel}+\partial_t
\bar{w}_{E_\perp}|$, does vary somewhat as a function of spatial
location $\V{r}_0$. The mean and standard deviation of this
parallel-to-total energy transfer ratio---averaged over the entire
simulation time $T$ and over all 64 diagnosed points $\V{r}_0$---is equal
to $0.67\pm 0.24$, with no significant variation for different choices
of $\tau$. This result indicates that both $E_\parallel$ and $E_\perp$
contribute to the energy transfer, though points where one component
dominates over the other will be highlighted in the following
sections.  


\section{Comparing Electric Field and Heat Flux contributions}
\label{sec:heatflux}

As shown by Eqn.~\ref{eqn:dws}, the change of phase-space energy
density at a given point in 3D-3V phase space $(\V{r},\V{v})$ is the
sum of changes due to each of the three terms on the right-hand side
of the equation.  The first term represents the advective heat flux,
the second is the work done on the particles by the total electric
field (the sum of work done by $E_\parallel$ and $E_\perp$), and the
third is the work done by the magnetic field (which must be zero when
integrated over velocity space). The field-particle correlation
provides information about the work done by the electric field, but it
is worthwhile to analyze how all of the terms lead to the net energy
transfer to or from the protons at a single point in space.

To calculate the net rate of change of the spatial energy density at a
single point $\V{r}_0$ in time, $\partial_t\bar{w}(\V{r}_0,t)$, we may
simply integrate Eqn.~\ref{eqn:dws} over all velocity space,
identifying each of the different terms.  Note that this equation must
be satisfied instantaneously, so we do not time-average the
correlations in this analysis. The total rate of  change in the proton spatial
energy density is given by
\begin{equation}
  \frac{\partial\bar{w}}{\partial t}(\V{r}_0,t)=\frac{\partial}{\partial t}\int d\V{v}\frac{m v^2 f}{2}.  
\end{equation}
 The instantaneous rate of
work done by the parallel and perpendicular components of the electric
field on the protons is simply given by Eqn.~\ref{eqn:partialw} with a
correlation interval $\tau=0$. Note that $\tau=0$ will be implicitly
assumed unless otherwise mentioned for all determinations of
$\partial_t\bar{w}_{E_\parallel}(\V{r}_0,t)$ and $\partial_t
\bar{w}_{E_\perp}(\V{r}_0,t)$ for the remainder of this section, and
the total rate of work done by the electric field is given by
$\partial_t \bar{w}_{E} (\V{r}_0,t) =
\partial_t\bar{w}_{E_\parallel}(\V{r}_0,t) + \partial_t
\bar{w}_{E_\perp}(\V{r}_0,t)$.  Following this procedure, the rate of
change of the proton spatial energy density due to the magnetic field
is given by
\begin{equation}
  \partial_t \bar{w}_{B} (\V{r}_0,t)=\frac{q_s}{c} \int d\V{v}
  \frac{v^2}{2} \left(\mathbf{v} \times \mathbf{B}\right) \cdot
  \frac{\partial f_s}{\partial \mathbf{v}}.
  \label{eqn:partialB}
\end{equation}
\change{As with the energy transfer rates calculated in
  Section~\ref{sec:fpem}, all of the rates in this section are
  calculated in the time-averaged bulk-velocity frame for each spatial
  point in the simulation.} Note that integrating by parts in velocity
of Eqn.~\ref{eqn:partialB} enables the integrand to be manipulated
into the form $\mathbf{v} \cdot \left(\mathbf{v} \times
\mathbf{B}\right)f_s =0$, so the net work done by the magnetic field
must equal zero, as expected.  Nonetheless, evaluating
Eqn.~\ref{eqn:partialB} with the numerical velocity derivatives
provides a convenient means for estimating the accuracy of the
integration and assessment of $\partial_{\V{v}}f_s$.

Since the spatial gradients in the ballistic (advective) term in
Eqn.~\ref{eqn:dws} are not available with only single-point
measurements, we cannot directly evaluate this term.  But, since we
can determine all of the other terms in the equation \change{using
  single-point measurements,\footnote{\change{As discussed in Section
      7.1 of \cite{Howes:2017a}, whether our measurement of the plasma
      occurs at a single point, or along a single trajectory, it is
      sufficient for the correlation to be averaged over an interval
      longer than $2\pi$ of the phase of the wave, $\phi=\mathbf{k}
      \cdot \mathbf{v} - \omega t$, in order to resolve the nature of
      the secular transfer of energy.}}} we may obtain the
contribution from the advective heat flux at point $\V{r}_0$ by
combining all of the other terms,
\begin{equation}
\partial_t \bar{w}_{\textrm{Ball}}(\V{r}_0,t)=\partial_t \bar{w}
-\partial_t \bar{w}_{E} (\V{r}_0,t)-\partial_t \bar{w}_{B}
(\V{r}_0,t).
\label{eqn:partialBall}
\end{equation}

\begin{figure}
  \centerline{\includegraphics[width = 14cm, viewport = 10 0 370 350,
      clip=true]
    {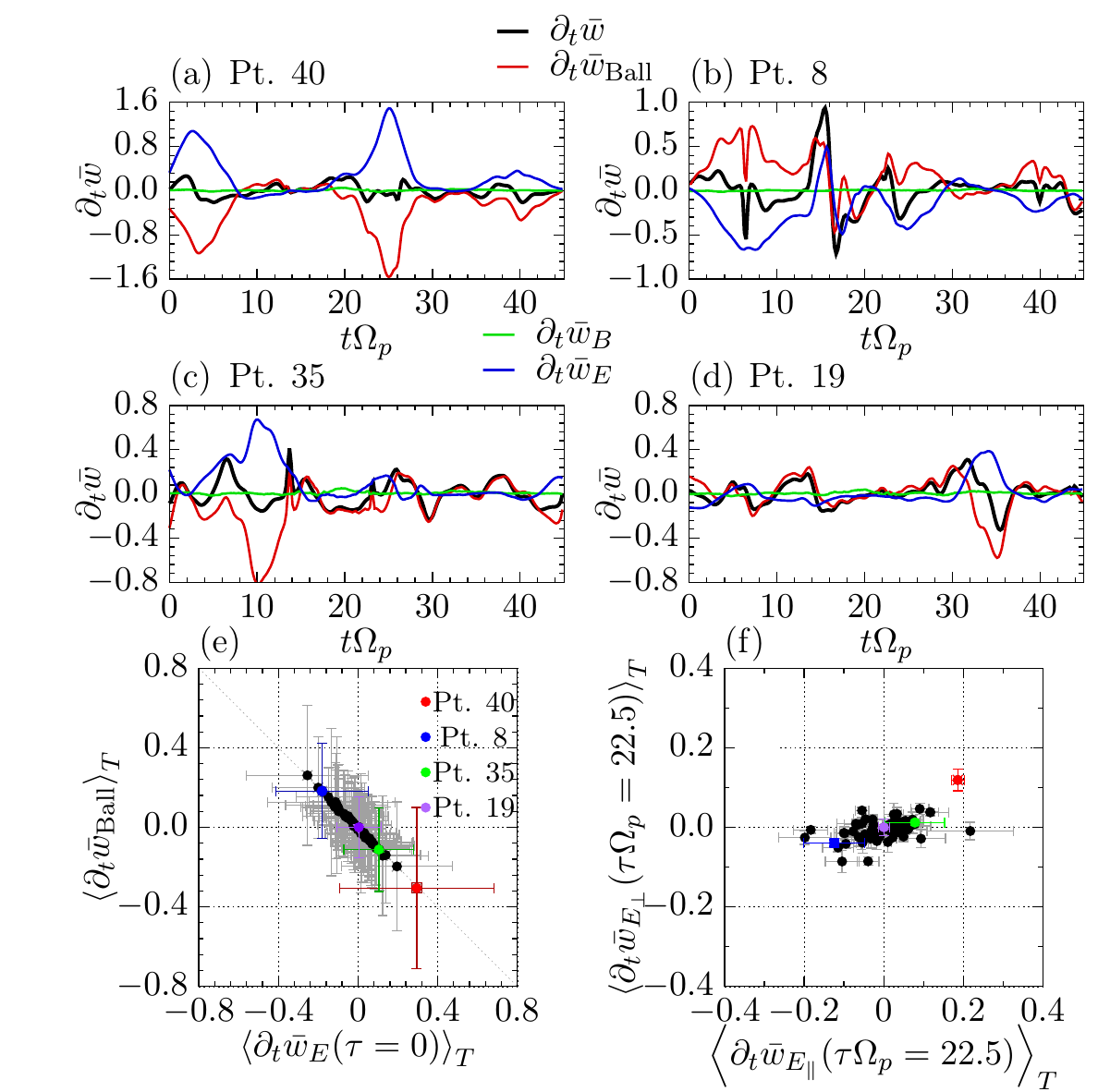}}
  \caption{(a-d) Components of the energy density transfer rate
    $\partial_t\bar{w}$ due to the electric field (blue), magnetic
    field (green), and the ballistic term (red), as well as the
    overall transfer rate (black) at four points in the simulation
    domain. (e) The mean and standard deviation of the ballistic and
    instantaneous (correlation length $\tau=0$) electric field
    transfer rates at each spatial point. (f) The mean and standard
    deviation of the transfer rate due to the perpendicular and
    parallel electric field, with correlation length
    $\tau\Omega_p=22.5$.}
\label{fig:stat}
\end{figure}

In Figure~\ref{fig:stat}, we plot the time evolution of the
contributions to the rate of change in the proton spatial energy
density at spatial points (a) 40, (b) 8, (c) 35, and (d) 11: (i) total
change in proton spatial energy density $\partial_t
\bar{w}(\V{r}_0,t)$ (black); (ii) the ballistic (heat flux)
contribution $\partial_t \bar{w}_{\textrm{Ball}}(\V{r}_0,t)$, (red);
(iii) the total electric field contribution $\partial_t
\bar{w}_{E}(\V{r}_0,t)$ (blue); and (iv) the magnetic field
contribution $\partial_t \bar{w}_{B}(\V{r}_0,t)$ (green).  As expected
the contribution from the magnetic field is nearly zero, providing a
practical diagnostic for the accuracy of our velocity derivatives of
the 3V distribution function at a given point, $f_p(\V{r}_0,\V{v},t)$.

A salient, and somewhat unexpected, feature that stands out in the
time series in panels (a)--(d) is the strong anti-correlation of the
ballistic (red) and electric field (blue) terms. This anti-correlation
can be quantified by plotting the mean value, over the entire
simulation duration $T$, of the energy transfer due to these two
terms, $\langle \partial_t \bar{w}_{\textrm{Ball}} \rangle_T$ and
$\langle \partial_t \bar{w}_E\rangle_T$ against each other for each of
the 64 spatial points, as shown in Figure~\ref{fig:stat}(e) with error
bars given by the standard deviations of the means.  While the
standard deviation of these energy transfer rates has a significant
spread, the mean values are well described with a linear fit of
$\partial_t \bar{w}_{\textrm{Ball}}=-1.005\partial_t
\bar{w}_{E}+0.004$, a nearly perfect anti-correlation.

One plausible interpretation of this finding is that, when energy is
transferred to the protons by the electric field, the heat flux
efficiently advects the energy away. Conversely, when energy is lost
from the protons through energy transfer to the electric field, the
heat flux causes a net energy flow to that point in space.  In other
words, energy is efficiently transported away from regions where
$\V{j}\cdot\V{E}$ is positive, and toward regions where
$\V{j}\cdot\V{E}$ is negative.  The anti-correlation between
$\partial_t \bar{w}_E$ and $\partial_t \bar{w}_{\textrm{Ball}}$
supports the general picture of two different recent energy transport
models, described in \cite{Yang:2017} and \cite{Howes:2018a}, where
the electromagnetic work done on the particles is merely one step
in a process of converting turbulent energy into plasma heat. It must
be emphasized that, although locally the field-particle and advective
terms nearly cancel out, only the field-particle term represents a net
(integrated over configuration space volume) change in the particle
energy, and thus represents particle energization.  The ballistic term
simply leads to a transport in configuration space of the energy
gained by the particles when $\V{E}$ does work.

The anti-correlation identified in Figure~\ref{fig:stat}(e) does not
establish cause and effect: is the change in spatial energy density
driven by the work done by the electric field or by the heat flux?  A
detailed look at the time evolution of the heat flux and electric
field terms over the time range $0 \le t \Omega_p \le 8$ in
Figure~\ref{fig:stat}(a) suggests that the work by the electric field
is the primary driver of the energy evolution.  Over the interval $0
\le t \Omega_p \le 3$, the electric field is energizing the protons
(blue), and the rate of change of the spatial energy density is positive
(black).  At $t \Omega_p =3$, the rate of work done by the electric
fields peaks and then begins to decline; at the same time, the rate of
change spatial energy density swings to negative (black), suggesting
that the removal of energy density by the heat flux (red) begins to
dominate, advecting energy away from the diagnosed point.  This
evolution suggests that first the electric field energizes the protons
locally, and subsequently the extra energy is carried away by
advection.

Finally, in Figure~\ref{fig:stat}(f), we plot the time-averaged rate
of work done by the parallel electric field $\langle
\partial_t\bar{w}_{E_\parallel}(\V{r}_0,t,\tau\Omega_p=22.5)\rangle_T$
and perpendicular electric field $\langle \partial_t
\bar{w}_{E_\perp}(\V{r}_0,t,\tau\Omega_p=22.5) \rangle_T $ against one
another, with error bars from the standard deviations.  We find that
the statistical correlation between parallel and perpendicular
energization is fairly weak, with a mean and standard deviation of
$0.15 \pm 0.53$, indicating that the energy transfer due to
$E_\parallel$ and $E_\perp$ are not strongly correlated.

In summary, the unexpectedly clear anti-correlation found here between
the heat flux and electric field terms motivates a more detailed
investigation of their time evolution, with an aim to identify cause
and effect, rather than just anti-correlation. Consideration of the
contribution of the heat flux to the rate of change of the spatial
energy density, especially in systems with significant spatial
inhomogeneities, will be essential for fully characterizing the entire
chain of energy transport from turbulent plasma flows and
electromagnetic fields to plasma heat.


\section{Conclusions}
\label{sec:conclude}
We present in this work the first application of the field-particle
correlation technique to a system of \Alfven Ion-Cyclotron turbulence,
using electromagnetic field and proton distribution data drawn from an
\texttt{HVM} numerical simulation of kinetic protons and fluid
electrons. Unlike previous tests of the field-particle correlation
technique using gyrokinetic simulations of strong plasma turbulence
that prohibit the possibility of ion cyclotron damping
\citep{Klein:2017b}, the use of a hybrid code enables collisionless
energy transfer to the protons via both the Landau and cyclotron
resonances.  An isotropic simulation domain over a range of
wavevectors spanning ion kinetic scale lengths was chosen here to
allow proton energization by both Landau damping and cyclotron
damping. \changetwo{This simulation domain is not necessarily
  representative of solar wind turbulence, which is typically found to
  have more significant wavevector anisotropies.}

The first key finding of this study is that we have provided the first
numerical determination of the characteristic velocity-space signature
of proton cyclotron damping in a strong turbulence simulation using
the field-particle correlation technique, shown in
Fig.~\ref{fig:vspacesig}(a). The region of velocity space controlling
the energy transfer---$1 \le v_\perp/v_{tp} \le 3$ and $-2 \le
v_\parallel/v_{tp} \le 2$---is largely consistent with the formation
of a cyclotron diffusion plateau, of the kind observed in \emph{in
  situ} solar wind measurements, \emph{e.g.} \cite{He:2015}. The
velocity region of energization is inconsistent with the predictions
of stochastic heating by low-frequency \Alfvenic turbulence
\citep{Chandran:2010a}, which is predicted to preferentially heat
particles with $v_\perp/v_{tp} \lesssim 1$ \citep{Klein:2016a}.

Our study also confirmed the characteristic bipolar velocity-space
signature of Landau damping with an independent numerical code,
confirming previous determinations in single kinetic \Alfven wave
simulations \citep{Howes:2017b,Klein:2017b}, gyrokinetic simulations
of strong plasma turbulence \citep{Klein:2017b}, and observations of
the Earth's turbulent magnetosheath \citep{Chen:2019}. The
determination of the velocity-space signatures of both cyclotron
damping and Landau damping acting simultaneously at the same point
clearly demonstrates a second key result: the field-particle
correlation method can successfully employ single-point measurements
to distinguish distinct mechanisms of energy transfer occurring at the
same point in space.  Note that, although a simple decomposition of
the components of $\V{j}\cdot\V{E}$ can separate the perpendicular and
parallel contributions, the velocity-space signatures generated by the
field-particle correlation technique provide a practical means to
identify definitively the physical mechanisms that are responsible,
even when multiple channels of energization are occurring
simultaneously.

This study also quantitatively characterized the variations of the
velocity-space signatures of proton cyclotron damping and Landau
damping at different points in space and time, finding that the
pattern of energy transfer in velocity space generally persists,
although the signs of the energy transfer can switch since
collisionless wave-particle interactions are reversible, sometimes
leading to energy transfer from the particles to the electric field.

An unexpected finding here is a strong anti-correlation of the rate of
change of spatial energy density at a single point between the
ballistic (advective heat flux) and electric field terms. Preliminary
indications suggest that, first, the electric field energizes the
protons locally, and subsequently the extra energy is carried away by
advection, but a more detailed investigation of the time evolution of
these physical mechanisms that change the local spatial energy density
is required to confirm this hypothesis.

Further work is needed to explore the variation of the velocity-space
signatures of different particle energization mechanisms with changes
in the plasma parameters (\emph{e.g.}, $\beta_p$ and $T_p/T_e$) and
the characteristics of the turbulence (\emph{e.g.}, nonlinear
parameter $\chi$ and anisotropy of turbulence in wavevector
space). Ultimately, we aim to develop a framework of characteristic
velocity-space signatures of different proposed particle energization
mechanisms using the field-particle correlation technique, which
unlike other methods for studying plasma heating and particle
energization that require measurements of spatial gradients
(\emph{e.g.} \cite{Yang:2017}) is designed to be implemented using only
single-point measurements. This framework can then be used to
interpret the results of the field-particle correlation analysis of
single-point particle velocity distribution and electromagnetic field
measurements from current and future spacecraft missions, such as
\emph{Magnetospheric MultiScale} and \emph{Parker Solar Probe}. The
ultimate goal is to identify the dominant mechanisms of particle
energization and compute the resulting rates of particle energization
due to the damping of turbulence in key regions of the
heliosphere---the solar corona, solar wind, and planetary
magnetospheres.



The authors would like to thank Chris Chen, Justin Kasper, Matt Kunz,
and Lev Arzamasskiy for helpful discussions during the execution of
this project. K.G.K.~was supported by NASA grants 80NSSC19K1390 and
80NSSC19K0912. J.M.T.~was supported by NSF SHINE award AGS-1622306, and
G.G.H.~was supported by NASA grants HSR 80NSSC18K1217, HGI
80NSSC18K0643, and MMSGI 80NSSC18K1371. F.V. has been partially
supported by the Agenzia Spaziale Italiana under the contract
No. ASI-INAF 2015-0390R.O Numerical simulations have been performed on
the supercomputer MARCONI at CINECA, Italy.



\appendix

\section{\change{Single-Point Mode Identification}}
\label{app:modes}

In this paper, we present the velocity-space signature of ion
cyclotron damping using the field-particle correlation technique,
illustrated in Fig 4 (a).  To establish that this is indeed due to the
ion cyclotron resonance, we show here that the simulation of
turbulence indeed contains ion cyclotron waves that are expected to
damp collisionlessly via the ion cyclotron resonance.

A common method to diagnose the nature of simulated turbulence is to
calculate power as a function of both frequency and length scale, and
compare the result to linear predictions, producing so-called
$\omega-k$ diagrams. Such diagrams are not necessarily a reliable way
to identify wave modes in strong plasma turbulence.  For example, in
Fig 5 of \cite{TenBarge:2012a}, a plot of $\omega$ vs. $k_\perp$ for
strong KAW turbulence shows significant broadening, which is
interpreted to be due to the strong nonlinear energy transfer among
modes, and is not directly comparable to the typical linear $\omega(\V{k})$
dispersion relations.

To identify the nature of the turbulence simulated in this work using
the single-point time series presented in the main text, we consider
the relations among different components of the turbulent fluctuations
and compare to the predicted eigenfunctions for different wave modes
from linear kinetic theory. The practice of calculating these
relations, including various helicities, polarizations and other
transport ratios
\citep{Gary:1986,Gary:1992,Gary:1993,Song:1994,Krauss-Varban:1994},
has a long history of application to in situ observations of both the
magnetosphere \citep{Lacombe:1995,Denton:1995,Schwartz:1996,Zhu:2019}
and solar wind
\citep{He:2011a,Salem:2012,TenBarge:2012b,Chen:2013a,Roberts:2013,Klein:2014a,Verscharen:2017,Wu:2019};
see \cite{Klein:Thesis:2013} for a more exhaustive review.

  \begin{figure}
  \centerline{\includegraphics[width = 14cm]
    {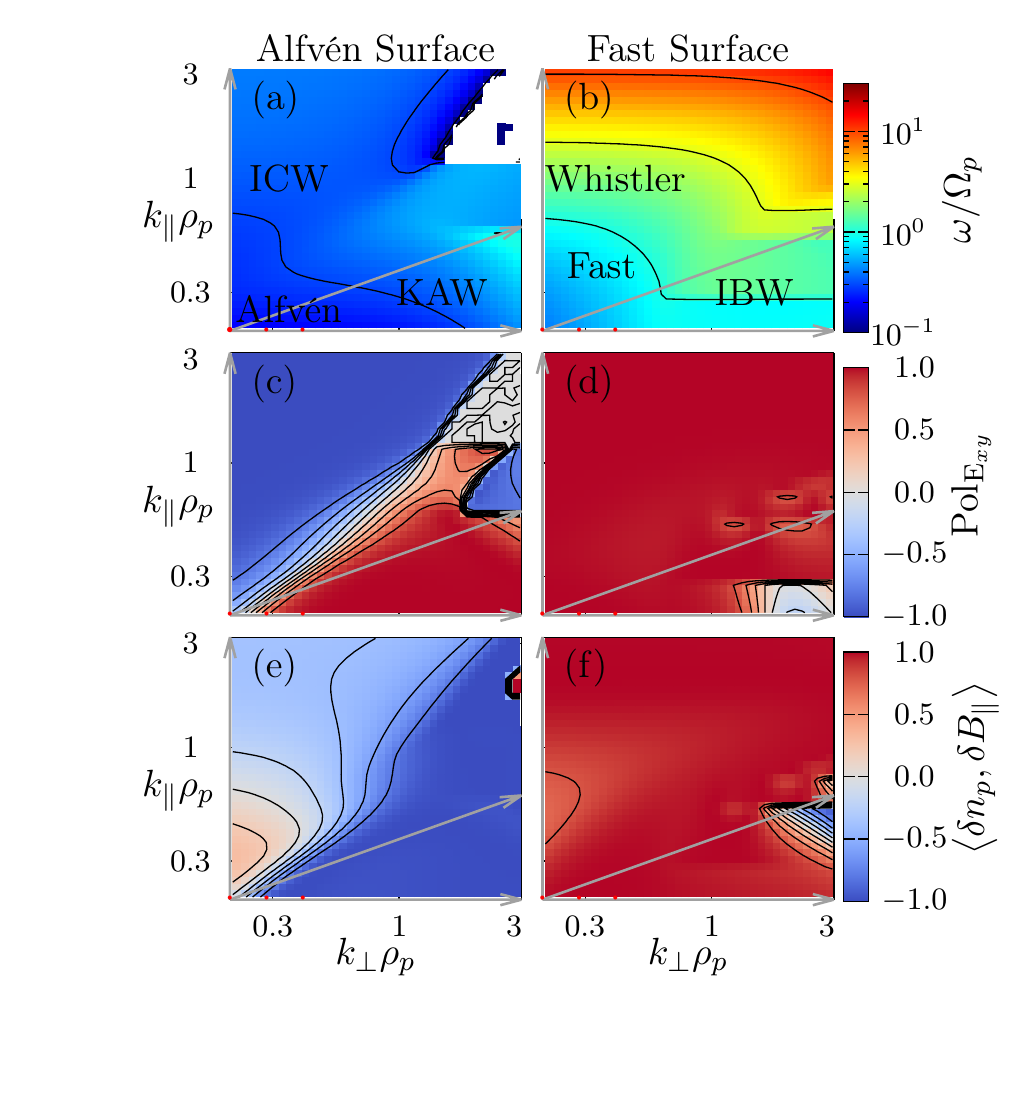}}%
  \caption{(a,b) Normalized frequency $\omega(\V{k}\rho_p)/\Omega_p$,
    (c,d) electric field polarization Pol$_{E_{xy}}(\V{k}\rho_p)$,
    Eqn.~\ref{eq:Epol}, and (e,f) density-magnetic field correlation
    $\left<\delta n, \delta B_\parallel\right>(\V{k}\rho_p)$, Eqn.~\ref{eq:CnB}, for
    the \Alfven and fast dispersion surfaces over wavevectors
    simulated in this work. The grey lines indicate parallel, oblique,
    and perpendicular cuts used for comparison to frequency series in
    Fig.~\ref{fig:pol_one}.}
\label{fig:freq_one}
\end{figure}

Two particularly useful measures to distinguish between the normal
modes accessible to the region of wavevector space simulated in this
work, namely ion cyclotron (ICW) and kinetic Alfv\'en (KAW) waves on the Alfv\'en
dispersion surface, and whistlers on the fast dispersion surface, are
the circular polarization of the electric field about the magnetic
field,
\begin{equation}
  \textrm{Pol}_{E_{xy}}=\frac{i(E_xE_y^*-E_x^*E_y)}{|E_x||E_y|}
  \label{eq:Epol}
\end{equation}
and the density-magnetic field correlation
\citep{Howes:2012a,Klein:2012},
  \begin{equation}
    \left<\delta n, \delta B_\parallel\right> =\frac{
      \left(\delta n^* \delta B_\parallel+\delta n \delta B_\parallel^* \right)}
    {|\delta n||\delta B_\parallel|}
    \label{eq:CnB}, 
  \end{equation}
  where $\delta \V{E}$, $\delta \V{B}$, and $\delta n$ are
  complex-valued Fourier coefficientss. These two eigenfunction
  relations, along with the normal mode frequencies $\omega/\Omega_p$,
  for the \Alfven and fast dispersion surfaces are plotted in
  Fig.~\ref{fig:freq_one}.
  
  The electric field polarization changes sign between the parallel
  and perpendicular kinetic extensions of the \Alfven solution, from
  left-handed ICWs to right-handed KAWs. The fast modes are nearly
  uniformly right-handed over this wavevector regime. Density and
  magnetic field fluctuations are strongly anti-correlated for oblique
  \Alfven solutions, weakly correlated for parallel \Alfven solutions,
  and strongly correlated for all fast mode solutions.

  \begin{figure}
    \centerline{\includegraphics[width = 14cm]
      {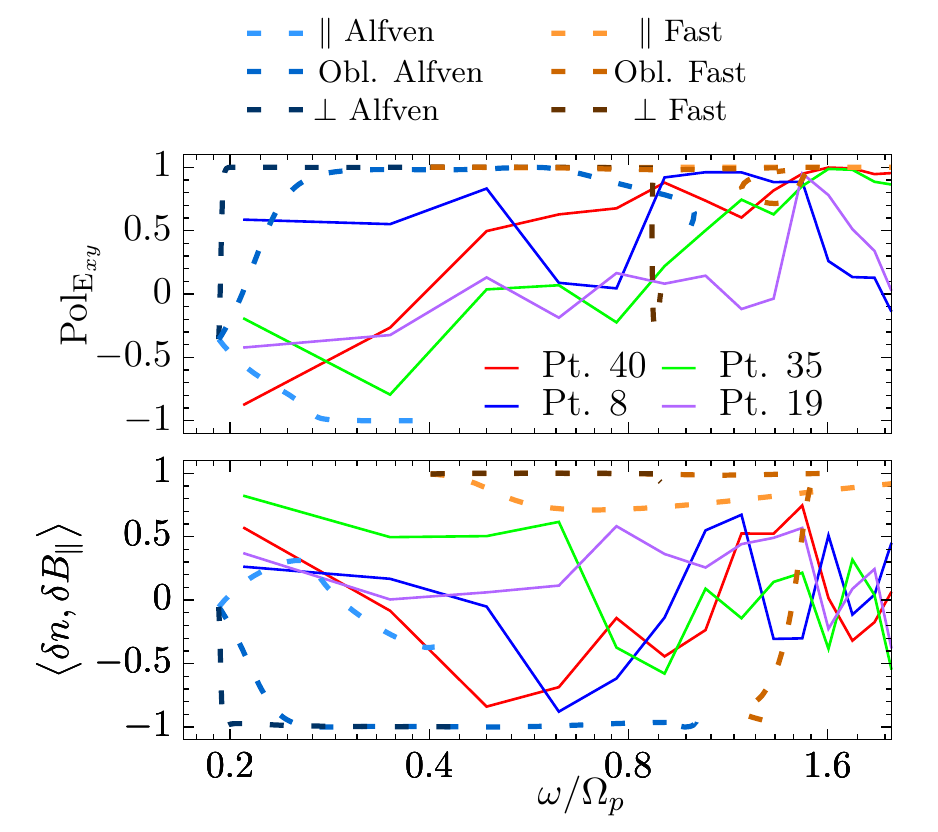}}%
    \caption{Electric field polarization, Eqn.~\ref{eq:Epol}, and
      density-magnetic field correlation, Eqn.~\ref{eq:CnB},
      calculated from frequency spectra drawn from the four spatial
      points investigated in this work. Expectations from linear
      theory along the grey arrows in Fig.~\ref{fig:freq_one} for both
      the \Alfven and fast dispersion surfaces are indicated with
      dashed lines.}
    \label{fig:pol_one}
  \end{figure}

These two eigenfunction relations can be used to identify the presence
of ICWs in our turbulent simulation using only single-point time
series of measurements, similar to what is measurable with spacecraft
missions.  At each of the four spatial points examined in the main
text, we Fourier transform in time to obtain the complex Fourier
coefficients (as a function of frequency) for $E_x$, $E_y$, $\delta
n$, and $\delta B_\parallel$.  From these complex Fourier
coefficients, we compute the circular polarization using
\eqref{eq:Epol} and the density-magnetic field correlation using
\eqref{eq:CnB} as a function of normalized angular frequency
$\omega/\Omega_p$ (solid lines in Fig.~\ref{fig:pol_one}).

To compare to the predicted variation of these eigenfunction relations
for the waves from linear kinetic theory, we compute the values
Pol$_{E_{xy}}$ and $\left<\delta n, \delta B_\parallel\right>$ along
particular trajectories through $(k_\perp,k_\parallel)$ wavevector
space, indicated in Fig.~\ref{fig:freq_one} by the gray
arrows \footnote{Note that the values of Pol$_{E_{xy}}$ from the
  second row and $\left<\delta n, \delta B_\parallel\right>$ from the
  third row are plotted against the corresponding frequency
  $\omega/\Omega_p$ from the first row.}.  For example, the
``parallel'' path (vertical gray arrow) transitions from the regime of
\Alfven waves to the regime of ICWs, whereas the ``perpendicular'' path
(horizontal gray arrow) transitions from the regime of \Alfven waves
to the regime of KAWs. These predicted theoretical values are plotted in  
 Fig.~\ref{fig:pol_one} as dashed lines.

Examining first Pol$_{E_{xy}}$ in Fig.~\ref{fig:pol_one}, at the
lowest frequencies $\omega/\Omega_p \le 0.4$ (which correspond only to
the \Alfven solutions, as all of the fast wave modes have higher
frequencies $\omega/\Omega_p \ge 0.4$), we find Pol$_{E_{xy}}<0$ for
three of the four spatial points. The only region for the \Alfven or
fast solutions that has Pol$_{E_{xy}}<0$ is the ICW regime, so we can
conclude that, at those three points, there exists a significant
contribution of ICW fluctuations.  At $\omega/\Omega_p > 0.4$, we find
Pol$_{E_{xy}} \ge 0$ at all four points, suggesting that the
fluctuations at these frequencies are either KAWs or any of the fast
mode fluctuations.

Turning next to $\left<\delta n, \delta B_\parallel\right>$ in
Fig.~\ref{fig:pol_one}, again at the lowest frequencies
$\omega/\Omega_p \le 0.4$ we find a $\left<\delta n, \delta
B_\parallel\right> > 0$, agreeing well with the prediction for ICWs
(light blue, dashed line).  Shifting to the frequency range $0.4 \le
\omega/\Omega_p \le 0.9$, we find $\left<\delta n, \delta
B_\parallel\right> < 0$ for three of the four spatial points.  Since
only the KAW regime has $\left<\delta n, \delta B_\parallel\right> <
0$, we conclude that a substantial fraction of the fluctuations in
this frequency range are KAWs.

In conclusion, at the lowest frequencies $\omega/\Omega_p \le 0.4$,
the combination of Pol$_{E_{xy}}<0$ and $\left<\delta n, \delta
B_\parallel\right> > 0$ provides strong evidence that we indeed
observe ICWs in our turbulence simulation.  Furthermore, looking at
magnetic and electric frequency power spectra in
Fig.~\ref{fig:spectra}, there is significant power at these low
frequencies, so we expect that ion cyclotron damping may indeed play a
key role in the removal of energy from the turbulent fluctuations in
the simulation.  In addition, in the frequency range $0.4 \le
\omega/\Omega_p \le 0.9$, the combination of Pol$_{E_{xy}}>0$ and
$\left<\delta n, \delta B_\parallel\right> < 0$ provides strong
evidence for the presence of KAWs in the turbulence
simulation. Therefore, we may expect to see signatures of ion Landau
damping in our simulation.
   
\bibliographystyle{jpp}


\end{document}